\documentclass[11pt,draftclsnofoot,onecolumn]{IEEEtran}

\input{mysymbol.sty}
\usepackage{graphicx,slashbox,dsfont}
\usepackage{epsfig}
\usepackage{amsmath}
\usepackage{bbm}
\usepackage{amssymb}
\usepackage{subfigure}
\usepackage{wrapfig}
\usepackage{times,color}
\usepackage{cite,footnote,xspace,syntonly,algorithm,algorithmic,bm}
\usepackage{algorithm,algorithmic}
\usepackage[caption=false,font=footnotesize]{subfig}

\newcommand{\calN}{{\cal N}}

\def \cP {\mathcal{P}}

\def \bY {{\bf Y}}
\def \bR {{\bf R}}
\def \bX {{\bf X}}
\def \bA {{\bf A}}
\def \bV {{\bf V}}

\def \bF {{\bf F}}
\def \bG {{\bf G}}
\def \bI {{\bf I}}

\def \bM {{\bf M}}
\def \bU {{\bf U}}
\def \bV {{\bf V}}

\def \bD {{\bf D}}
\def \bZ {{\bf Z}}

\def \bH {{\bf H}}
\def \bL {{\bf P}}
\def \bQ {{\bf Q}}

\def \bw {{\bf w}}

\def \bv {{\bf v}}
\def \bx {{\bf x}}
\def \bz {{\bf z}}

\def \be {{\bf e}}
\def \br {{\bf r}}

\def \bv {{\bf v}}
\def \ba {{\bf a}}
\def \bl {{\bf p}}
\def \bq {{\bf q}}
\def \by {{\bf y}}
\def \bs {{\bf s}}

\def \bPhi {{\bf \Phi}}

\def \ind {\mathds{1}}

\def \tr {\text{tr}}

\long\def\symbolfootnote[#1]#2{\begingroup
\def\thefootnote{\fnsymbol{footnote}}
\footnote[#1]{#2}\endgroup} \psfull

\begin{document}


\title{\huge Dynamic Anomalography: Tracking Network\\
Anomalies via Sparsity and Low Rank$^\dag$}

\author{{\it Morteza Mardani, Gonzalo~Mateos, and Georgios~B.~Giannakis~(contact author)$^\ast$}}

\markboth{IEEE JOURNAL OF SELECTED TOPICS IN SIGNAL PROCESSING (SUBMITTED)}
\maketitle \maketitle \symbolfootnote[0]{$\dag$ Work in this
paper was supported by the MURI Grant No. AFOSR FA9550-10-1-0567.
Parts of the paper appeared in the {\it Proc. of
the 45th Asilomar Conference on Signals, Systems, and
Computers}, Pacific Grove, CA, Nov. 6-9, 2011.} 
\symbolfootnote[0]{$\ast$ The authors are with the Dept.
of Electrical and Computer Engineering, University of
Minnesota, 200 Union Street SE, Minneapolis, MN 55455. Tel/fax:
(612)626-7781/625-4583; Emails:
\texttt{\{morteza,mate0058,georgios\}@umn.edu}}

\vspace*{-80pt}
\begin{center}
\small{\bf Submitted: }\today\\
\end{center}
\vspace*{10pt}


\thispagestyle{empty}\addtocounter{page}{-1}
\begin{abstract}
In the backbone of large-scale networks, origin-to-destination (OD) 
traffic flows experience abrupt unusual changes known
as \textit{traffic volume anomalies}, which can result in congestion and 
limit the extent to which end-user quality of service requirements are met. 
As a means of maintaining seamless end-user experience in dynamic environments, 
as well as for ensuring network security, this paper deals with a crucial 
network monitoring task termed \textit{dynamic anomalography}. 
Given link traffic measurements (noisy superpositions of
unobserved OD flows) periodically acquired by backbone routers, 
the goal is to construct an estimated \textit{map} of anomalies in \textit{real time}, 
and thus summarize the network `health state' along both the flow and time dimensions.
Leveraging the low intrinsic-dimensionality of OD flows and the
\textit{sparse} nature of anomalies, a novel online estimator is proposed
based on an exponentially-weighted least-squares criterion regularized with the 
sparsity-promoting $\ell_1$-norm of the anomalies, 
and the nuclear norm of the nominal traffic matrix. After recasting the 
non-separable nuclear norm into a form amenable to online
optimization, a real-time algorithm for dynamic anomalography is developed
and its convergence established under simplifying technical assumptions. 
For operational conditions where computational complexity reductions are at a premium, a lightweight
stochastic gradient algorithm based on Nesterov's acceleration
technique is developed as well. Comprehensive numerical 
tests with both synthetic and real network data corroborate the 
effectiveness of the proposed online algorithms and their tracking
capabilities, and demonstrate that they outperform state-of-the-art approaches developed
to diagnose traffic anomalies.
\end{abstract}

\vspace*{-5pt}
\begin{keywords}
Traffic volume anomalies, online optimization, sparsity, network cartography, low rank.
\end{keywords}
%
\newpage


\section{Introduction}
\label{sec:intro}

Communication networks have evolved from specialized, research and tactical 
transmission systems to large-scale and highly complex interconnections of 
intelligent devices. Thus, ensuring compliance to service-level agreements and
quality-of-service (QoS) guarantees necessitates ground-breaking management and 
monitoring tools providing operators with a comprehensive and updated 
view of the network landscape. Situational awareness provided by such tools 
will be a key enabler for effective information dissemination, routing and congestion
control, network health management, and security assurance.

In the backbone of large-scale networks, origin-to-destination (OD) 
traffic flows experience abrupt unusual changes which can result 
in congestion, and limit QoS provisioning of the end users. These so-termed \emph{traffic
volume anomalies} could be due to unexpected failures in networking equipment, cyberattacks
(e.g., denial of service (DoS) attacks), or, intruders which hijack the network
services~\cite{MC03}. Unveiling such anomalies in a promptly manner is a crucial monitoring task
towards engineering network traffic. This is a challenging
task however, since the available data are usually high-dimensional,
noisy and possibly incomplete link-load measurements, which are the superposition of
\emph{unobservable} OD flows. Several studies have experimentally demonstrated 
the low intrinsic dimensionality of the nominal traffic subspace, 
that is, the intuitive \textit{low-rank} property of the traffic matrix in the 
absence of anomalies, which is mainly due to common temporal patterns across OD flows,
and periodic behavior across time~\cite{LPC04,zrwq09}.
Exploiting the low-rank structure of the anomaly-free traffic matrix,
a landmark principal component analysis (PCA)-based method was put forth in~\cite{LCD04}
to identify network anomalies; see also~\cite{distributed_pca_heor12} for a distributed
implementation. A limitation of the algorithm
in~\cite{LCD04} is that it cannot identify the anomalous flows. 
Most importantly,~\cite{LCD04} has not exploited the \textit{sparsity} 
of anomalies across flows and time -- anomalous traffic spikes are rare, and tend to
last for short periods of time relative to the measurement horizon.

Capitalizing on the low-rank property of the traffic matrix and the sparsity of
the anomalies, the fresh look advocated here permeates benefits from
rank minimization~\cite{candes_moisy_mc,CLMW09,CR08}, 
and compressive sampling~\cite{candes_tutorial,CT05}, to perform \textit{dynamic anomalography}. 
The aim is to construct a map of network \textit{anomalies} in {real time},
that offers a succinct depiction of the network `health state' across
both the flow and time dimensions (Section \ref{sec:model}). 
Special focus will be placed on devising online (adaptive) 
algorithms that are capable of efficiently processing link measurements and 
track network anomalies `on the fly'; see also 
\cite{multivariate_onlineanomaly_lakhina07} for a `model-free' approach
that relies on the kernel recursive LS (RLS) algorithm. Accordingly, the novel online estimator 
entails an exponentially-weighted least-squares (LS) cost regularized with the 
sparsity-promoting $\ell_1$-norm of the anomalies, and the nuclear norm of the nominal traffic matrix.
After recasting the non-separable nuclear norm into a form amenable to online
optimization (Section \ref{ssec:separable}), a real-time algorithm for dynamic anomalography is developed
in Section \ref{sec:online_alg} based on alternating minimization. Each time a new datum is acquired,
anomaly estimates are formed via the least-absolute shrinkage and selection
operator (Lasso), e.g,~\cite[p. 68]{elements_of_statistics}, and the low-rank nominal traffic
subspace is refined using RLS~\cite{Solo_Adaptive_Book}. 
Convergence analysis is provided under simplifying technical assumptions in 
Section \ref{subsec:cnvg_analysis}. For situations were reducing computational complexity is critical, an online 
stochastic gradient algorithm based on Nesterov's accelaration technique~\cite{fista,nesterov83} is developed as well 
(Section \ref{ssec:nesterov}). 
The possibility of implementing the anomaly trackers in a distributed
fashion is further outlined in Section \ref{ssec:distributed}, 
where several directions for future research are also delineated.

Extensive numerical tests involving both synthetic and real network data 
corroborate the effectiveness of the proposed algorithms in unveiling
network anomalies, as well as their tracking
capabilities when traffic routes are slowly time-varying, and the
network monitoring station acquires incomplete link traffic measurements (Section \ref{sec:sims}).
Different from~\cite{zggr05} which employs a two-step batch procedure to learn
the nominal traffic subspace first, and then unveil anomalies via $\ell_1$-norm minimization,
the approach here estimates both quantities jointly and attains better performance as
illustrated in Section \ref{ssec:perf_real}. 
Concluding remarks are given in Section \ref{sec:conc}, while 
most technical details relevant to the convergence proof in Section \ref{subsec:proof} 
are deferred to the Appendix.

\noindent{\it Notation}: Bold uppercase (lowercase) letters will denote
matrices (column vectors), and calligraphic letters will be used for sets.
Operators $(\cdot)'$, $\rm{tr}(\cdot)$, $\lambda_{\min}(\cdot)$, $[\cdot]_+$, and $\mathbb{E}[\cdot]$, 
will denote transposition, matrix trace, minimum eigenvalue, projection
onto the nonnegative orthant, and 
expectation, respectively; $|\cdot|$ will be used for the cardinality of a set, and the magnitude of a 
scalar. The positive semidefinite matrix $\mathbf{M}$ will be
denoted by $\bbM\succeq\mathbf{0}$.
The $\ell_p$-norm of $\bx \in \mathbb{R}^n$ is $\|\bx\|_p:=(\sum_{i=1}^n 
|x_i|^p)^{1/p}$
for $p \geq 1$. For two matrices $\bM,\bU \in \mathbb{R}^{n \times n}$,
$\langle \bM, \bU \rangle := \rm{tr(\bM' \bU)}$ denotes their trace inner 
product.
The Frobenious norm of matrix $\bM = [m_{i,j}] \in \mathbb{R}^{n \times p}$ is
$\|\bM\|_F:=\sqrt{\tr(\bM\bM')}$,
$\|\bM\|:=\max_{\|\bx\|_2=1} \|\bM\bx\|_2$ is the spectral norm,
$\|\bM\|_1:=\sum_{i,j} |m_{i,j}|$ is the $\ell_1$-norm, and 
$\|\bM\|_{\ast}:=\sum_{i}\sigma_i(\bM)$ is the nuclear norm, where
$\sigma_i(\bM)$ denotes the $i$-th singular value of $\bM$. The $n \times n$ 
identity matrix will be represented by $\bI_n$, while $\mathbf{0}_{n}$ will stand for 
the $n \times 1$ vector of all zeros, and $\mathbf{0}_{n \times p}:=\mathbf{0}_{n}
\mathbf{0}'_{p}$.


\section{Modeling Preliminaries and Problem Statement}
\label{sec:model}
Consider a backbone Internet protocol (IP) network naturally modeled as a 
directed graph $G(\cal{N},\cal{L})$, where $\cal{N}$ and $\mathcal{L}$ denote the sets of
nodes (routers) and physical links of cardinality $|\calN|=N$ and $|\mathcal{L}|=L$, respectively. 
The operational goal of the network is to transport a set of OD traffic flows $\cal{F}$ (with 
$|\mathcal{F}| = F$) associated with specific source-destination pairs. For backbone networks, the number
of network layer flows is much larger than the number of physical links $(F\gg L)$.
Single-path routing is adopted here, that is, a given flow's traffic is carried through multiple
links connecting the corresponding source-destination pair
along a single path. Let
$r_{l,f},\:l\in\mathcal{L}, f\in\mathcal{F}$, denote the flow
to link assignments (routing), which take the value one whenever flow $f$ is carried over
link $l$, and zero otherwise. Unless otherwise stated, 
the routing matrix $\bbR:=[r_{l,f}]\in\{0,1\}^{L\times F}$
is assumed fixed and given.  Likewise, let $z_{f,t}$ denote the unknown traffic
rate of flow $f$ at time $t$, measured in e.g., Mbps. At any given time instant $t$, the
traffic carried over link $l$ is then the superposition of the flow rates
routed through link $l$, i.e., $\sum_{f\in\mathcal{F}}r_{l,f}z_{f,t}$.

It is not uncommon for some of the flow rates to experience unusual abrupt changes. These
so-termed \textit{traffic volume anomalies} are typically due to unexpected network failures, or
cyberattacks (e.g., DoS attacks) which aim at compromising the
services offered by the network~\cite{MC03}.
Let $a_{f,t}$ denote the unknown traffic volume anomaly of flow $f$ at
time $t$. In the presence of anomalous flows, the measured traffic carried by link $l$
over a time horizon $t \in [1,T]$ is then given by
\begin{equation}
y_{l,t}=\sum_{f\in\cal{F}}r_{l,f}(z_{f,t}+a_{f,t}) + v_{l,t},~t=1,...,T \label{eq:y_lt}
\end{equation}
where the noise variables $v_{l,t}$ account for measurement errors
and unmodeled dynamics. 

In IP networks, traffic volume can be readily measured on a per-link basis using
off-the-shelf tools such as the simple network management protocol (SNMP)
supported by most routers. Missing entries in the
link-level measurements $y_{l,t}$ may however skew the network operator's perspective. SNMP packets may
be dropped for instance, if some links become congested, rendering link count information for those
links more important, as well as less available~\cite{Roughan}.
To model missing link measurements, collect the tuples $(l,t)$ associated with the
available observations $y_{l,t}$ in the set $\Omega \in [1,2,...,L] \times
[1,2,...,T]$. Introducing the matrices $\bY:=[y_{l,t}],
\bV:=[v_{l,t}]\in\mathbb{R}^{L \times T}$, and $\bZ:=[z_{f,t}], \bA:=[a_{f,t}]
\in\mathbb{R}^{F \times T}$, the (possibly incomplete) set of measurements in
\eqref{eq:y_lt}  can be expressed in compact matrix form as
\begin{equation}
\cP_{\Omega}(\bY)=\cP_{\Omega}(\bR \left(\bZ + \bA\right) + \bV) \label{eq:Y}
\end{equation}
where the sampling operator $\cP_{\Omega}(.)$ sets
the entries of its matrix argument not in $\Omega$ to zero, and keeps the rest
unchanged. Matrix $\bZ$ contains the nominal traffic flows over the time horizon of
interest. Common temporal patterns among the traffic
flows in addition to their periodic behavior, render most rows
(respectively columns) of $\bZ$ linearly dependent, and thus $\bZ$
typically has low rank. This intuitive property has been extensively validated
with real network data; see e.g,~\cite{LPC04}. Anomalies in $\bA$ are expected 
to occur sporadically over time, and last shortly
relative to the (possibly long) measurement interval $[1,T]$.
In addition, only a small fraction of the flows is  supposed to be anomalous at a any given
time instant. This renders the anomaly traffic matrix $\bA$ 
sparse across both rows (flows) and columns (time). 

Given measurements $\cP_{\Omega}(\bY)$
adhering to~\eqref{eq:Y} and the binary-valued routing matrix $\bR$, 
the main goal of this paper is to accurately estimate the anomaly matrix $\bA$, 
by capitalizing on the sparsity of $\bA$ and the low-rank property of $\bZ$. Special
focus will be placed on devising online (adaptive) algorithms that are capable
of efficiently processing link measurements and tracking network anomalies in real time.
This critical monitoring task is termed \textit{dynamic anomalography}, and 
the resultant estimated map $\hat{\bA}$ offers a depiction of the network's `health state' 
along both the flow and time dimensions.  If $|\hat{a}_{f,t}|>0$,
the $f$-th flow at time $t$ is deemed anomalous, otherwise it is healthy. By
examining $\bR$ the network operator can immediately determine the links carrying
the anomalous flows. Subsequently, planned contingency measures involving traffic-engineering algorithms
can be implemented to address network congestion.


\section{Unveiling Anomalies via Sparsity and Low Rank}
\label{sec:batch}
Consider the nominal link-count traffic matrix $\bX := \bR\bZ$, which inherits the low-rank property
from $\bZ$. Since the primary goal is to recover $\bA$, the following observation model 
\begin{equation}
\cP_{\Omega}(\bY)=\cP_{\Omega}(\bX + \bR \bA + \bV) \label{eq:Y_modf}
\end{equation}
can be adopted instead of \eqref{eq:Y}.
%
A natural estimator leveraging the low rank property of $\bX$ and 
the sparsity of $\bA$ will be sought next. The idea is to fit 
the incomplete data $\cP_{\Omega}(\bY)$ to the model $\bX + \bR \bA$ in 
the least-squares (LS) error sense, as well as minimize the
rank of $\bX$, and the number of nonzero entries of $\bA$ measured by its $\ell_0$-(pseudo)
norm. Unfortunately, albeit natural both rank and $\ell_0$-norm criteria
are in general NP-hard to optimize~\cite{Natarajan_NP_duro,rank_NP_Duro}.
Typically, the nuclear norm $\|\bX\|_*$ and the $\ell_1$-norm $\|\bA\|_1$
are adopted as surrogates, since they are the closest \textit{convex}
approximants to $\textrm{rank}(\bX)$ and $\|\bA\|_0$, respectively~\cite{RFP07,CT05,tropp06tit}.
Accordingly, one solves
\begin{equation}
\text{(P1)}~~~~~\min_{\{\bX,\bA\}} \frac{1}{2}\|\cP_{\Omega}(\bY - \bX -
\bR\bA)\|_{F}^{2} +\lambda_{\ast}\|\bX\|_{*} + \lambda_1\|\bA\|_1
\end{equation}
where $\lambda_*,\lambda_1\geq 0$ are rank- and sparsity-controlling parameters. 
When an estimate $\hat{\sigma}_v^2$ of the noise variance is available,
guidelines for selecting $\lambda_*$ and $\lambda_1$ have been proposed in~\cite{zlwcm10}.
Being convex (P1) is appealing, and it is known to attain good performance in
theory and practice~\cite{tit_exactrecovery_2012}. Also \eqref{eq:Y_modf}
and its estimator (P1) are quite general, as discussed in the ensuing remark.
\begin{remark}[Subsumed paradigms]\label{rem:sub_parad}
When there is no missing data and $\bX=\mathbf{0}_{L\times T}$, one is
left with an under-determined sparse signal recovery problem
typically encountered with compressive sampling (CS); see e.g.,~\cite{CT05} and the
tutorial account~\cite{candes_tutorial}. The
decomposition $\bY=\bX+\bA$ corresponds to principal component
pursuit (PCP), also referred to as robust principal component
analysis (PCA)~\cite{CLMW09,CSPW11}. PCP was adopted for
network anomaly detection using flow (not link traffic) measurements
in~\cite{ajwak10}. For the idealized
noise-free setting ($\bV=\mathbf{0}_{L\times T}$), sufficient conditions for exact recovery are
available for both of the aforementioned special cases~\cite{CT05,CLMW09,CSPW11}.
However, the superposition of a low-rank plus a \textit{compressed}
sparse matrix in \eqref{eq:Y_modf} further challenges identifiability
of $\{\bX,\bA\}$; see~\cite{tit_exactrecovery_2012} for early results. 
Going back to the CS paradigm, even when $\bX$ is nonzero one could envision a variant
where the measurements are corrupted with correlated (low-rank)
noise~\cite{Vaswani_Allerton_11}. Last but not least, when
$\bA=\mathbf{0}_{F\times T}$ and $\bY$ is noisy, the recovery of
$\bX$ subject to a rank constraint is nothing but PCA --
arguably, the workhorse of high-dimensional data
analytics. This same formulation is adopted for low-rank
matrix completion, to impute the missing entries of a low-rank
matrix observed in noise, i.e., $\cP_{\Omega}(\bY) =
\cP_{\Omega}(\bX +\bV)$~\cite{candes_moisy_mc}.
\end{remark}

Albeit convex, (P1) is a non-smooth optimization problem (both the nuclear and
$\ell_1$-norms are not differentiable at the origin). In addition, scalable algorithms
to unveil anomalies in large-scale networks should effectively overcome
the following challenges: (c1) the problem size can easily become
quite large, since the number of optimization variables is $(L+F)T$; (c2) 
existing iterative solvers for (P1) typically rely on costly SVD computations
per iteration; see e.g.,~\cite{tit_exactrecovery_2012}; and (c3)
different from the Frobenius and $\ell_1$-norms, (columnwise) 
nonseparability of the nuclear-norm challenges online processing
when new columns of $\cP_{\Omega}(\bY)$ arrive sequentially in time. In the remainder
of this section, the `big data' challenges (c1) and (c2) are dealt with to
arrive at an efficient batch algorithm for anomalography. 
Tracking network anomalies is the main subject of Section
\ref{sec:online_alg}.

To address (c1) and reduce the computational complexity and memory
storage requirements of the algorithms sought, it is henceforth assumed
that an upper bound $\rho \geq\textrm{rank}(\hat\bbX)$ 
is a priori available [$\hat{\bX}$ is the estimate obtained via (P1)].
As argued next, the smaller the value of $\rho$, the more efficient the
algorithm becomes. Small values of $\rho$ are well motivated due to the low
intrinsic dimensionality of network flows~\cite{LCD04}. Because $\textrm{rank}(\hat\bbX)\leq \rho$,
(P1)'s search space is effectively reduced and one can factorize the
decision variable as $\bX=\bL\bQ'$, where $\bL$ and $\bQ$ are
$L \times \rho$ and $T \times \rho$ matrices, respectively. It is possible to interpret 
the columns of $\bX$ (viewed as points in $\mathbb{R}^L$) as belonging to
a low-rank `nominal traffic subspace', spanned by the columns of $\bL$. The rows of $\bQ$
are thus the projections of the columns of $\bX$ onto the traffic subspace.

Adopting this reparametrization of $\bX$ in (P1), one arrives at
an equivalent optimization problem
\begin{align}
\text{(P2)}~~~~~\min_{\{\bL,\bQ,\bA\}}& \frac{1}{2}\|\cP_{\Omega}(\bY
- \bL\bQ' - \bR\bA)\|_{F}^{2} + \lambda_{\ast}\|\bL\bQ'\|_{\ast} +
\lambda_1 \|\bA\|_1 \nonumber
\end{align}
which is non-convex due to the bilinear terms $\bL\bQ'$. The number of 
variables is reduced from $(L+F)T$ in (P1),
to $\rho(L+T)+FT$ in (P2). The savings can be significant when $\rho$
is small, and both $L$ and $T$ are large.
Note that the dominant $FT$-term in the variable count of (P2) is due
to $\bbA$, which is sparse and can be efficiently handled even when
both $F$ and $T$ are large.


\subsection{A separable low-rank regularization}\label{ssec:separable}
To address (c2) [along with (c3) as it will become clear in
Section \ref{sec:online_alg}], consider the following alternative characterization
of the nuclear norm~\cite{RFP07,Recht_Parallel_2011}
\begin{equation}\label{eq:nuc_nrom_def}
\|\bX\|_*:=\min_{\{\bL,\bQ\}}~~~ \frac{1}{2}\left\{\|\bL\|_F^2+\|\bQ\|_F^2 \right\},\quad
\text{s. t.}~~~ \bX=\bL\bQ'.
\end{equation}
%
The optimization \eqref{eq:nuc_nrom_def} is over all possible bilinear factorizations
of $\bbX$, so that the number of columns $\rho$ of $\bL$ and $\mathbf{Q}$
is also a variable. Leveraging~\eqref{eq:nuc_nrom_def}, the following reformulation
of (P2) provides an important first step towards obtaining an online algorithm:
\begin{align}
\text{(P3)}~~~~~\min_{\{\bL,\bQ,\bA\}}& \frac{1}{2}\|\cP_{\Omega}(\bY
- \bL\bQ' - \bR\bA)\|_{F}^{2} +
\frac{\lambda_{\ast}}{2}\left\{\|\bL\|_F^2 + \|\bQ\|_F^2 \right\}+
\lambda_1\|\bA\|_1. \nonumber
\end{align}
As asserted in~\cite[Lemma 1]{tsp_rankminimization_2012}, adopting the
separable Frobenius-norm regularization in (P3) comes
with no loss of optimality relative to (P1), provided $\rho \geq\textrm{rank}(\hat\bbX)$.
By finding the global minimum of (P3)
[which could have considerably less variables than (P1)], one can  recover the optimal solution of
(P1). However, since (P3) is non-convex, it may have stationary points which need not
be globally optimum. Interestingly, the next proposition shows that under
relatively mild assumptions on $\textrm{rank}(\hat\bX)$
and the noise variance, every stationary point of (P3) is globally optimum for (P1).
For a proof, see~\cite[App. A]{tsp_rankminimization_2012}.

\begin{proposition}\label{prop:prop_1}
Let $\{\bar{\bL},\bar{\bQ},\bar{\bA}\}$ be a stationary point of (P3).
If $\|\cP_{\Omega}(\bY-\bar{\bL}\bar{\bQ}'-\bR\bar{\bA})\| \leq \lambda_*$,
then $\{\hat\bX:=\bar{\bL}\bar{\bQ}',\hat\bA=\bar{\bA}\}$ is the globally optimal solution of (P1).
\end{proposition}

\noindent The qualification condition $\|\cP_{\Omega}(\bY-\bar{\bL}\bar{\bQ}'-\bR\bar{\bA})\| \leq \lambda_*$
captures tacitly the role of $\rho$. In particular, for sufficiently small $\rho$ the residual
$\|\cP_{\Omega}(\bY-\bar{\bL}\bar{\bQ}'-\bR\bar{\bA})\| $
becomes large and consequently the condition is violated [unless $\lambda_*$ is large
enough, in which case a sufficiently low-rank solution to (P1) is expected]. The condition
on the residual also implicitly enforces $\textrm{rank}(\hat\bX) \leq \rho$, which is necessary for
the equivalence between (P1) and (P3). In addition, the noise 
variance affects the value of $\|\cP_{\Omega}(\bY-\bar{\bL}\bar{\bQ}'-\bR\bar{\bA})\|$,
and thus satisfaction of the said qualification inequality.


\subsection{Batch block coordinate-descent algorithm}\label{ssec:bcd_alg}

The block coordinate-descent (BCD) algorithm is adopted here 
to solve the batch non-convex optimization problem (P3). 
BCD is an iterative method which has been shown efficient in tackling 
large-scale optimization problems encountered with various
statistical inference tasks, see e.g.,~\cite{Bers}. The proposed solver entails an
iterative procedure comprising three steps per iteration $k = 1, 2,\ldots$

\begin{description}
\item [{\bf [S1]}]  \textbf{Update the anomaly map:}
    \begin{equation*}\bA[k+1]=
    \mbox{arg}\:\min_{\bA}\left[ \frac{1}{2} \|\cP_{\Omega}(\bY-\bL[k]\bQ'[k]-\bR\bA)\|_F^2 + \lambda_1 \|\bA\|_1  \right].
    \end{equation*}

\item [{\bf [S2]}]  \textbf{Update the nominal traffic subspace:}
        \begin{equation*}\bL[k+1]=
    \mbox{arg}\:\min_{\bL}\left[ \frac{1}{2}
    \|\cP_{\Omega}(\bY-\bL\bQ'[k]-\bR\bA[k+1])\|_F^2 +
    \frac{\lambda_{\ast}}{2} \|\bL\|_F^2  \right].
       \end{equation*}

\item [{\bf [S3]}] \textbf{Update the projection coefficients:}
            \begin{equation*}\bQ[k+1]=
    \mbox{arg}\:\min_{\bQ}\left[ \frac{1}{2}
    \|\cP_{\Omega}(\bY-\bL[k+1]\bQ'-\bR\bA[k+1])\|_F^2 +
    \frac{\lambda_{\ast}}{2} \|\bQ\|_F^2  \right].
       \end{equation*}
\end{description}
To update each of the variable groups, the cost of (P3) is minimized 
while fixing the rest of the variables to their most up-to-date
values. The minimization in [S1] decomposes over the columns of $\bA:=[\bba_1,\ldots,\bba_T]$. 
At iteration $k$, these columns are updated in parallel via
Lasso
\begin{equation}\label{eq:lasso}
\ba_t[k+1]=\arg\min_{\ba} \left[\frac{1}{2}
\|\mathbf{\Omega}_t(\by_t-\bL[k]\bq_t[k]-\bR\ba)\|_2^2 + \lambda_1\|\ba\|_1 \right],\quad t=1,\ldots,T
\end{equation}
where $\by_t$ and $\bq_t[k]$ respectively denote the $t$-th column of $\bY$ and $\bQ'[k]$,
while the diagonal matrix
$\bm{\Omega}_t \in \mathbb{R}^{L \times L}$ contains a one on its $l$-th
diagonal entry if $y_{l,t}$ is observed, and a zero otherwise. In practice, 
each iteration of the proposed algorithm minimizes \eqref{eq:lasso} inexactly, by performing
one pass of the cyclic coordinate-descent algorithm in~\cite[p. 92]{elements_of_statistics};
see Algorithm \ref{tab:table_1} for the detailed iterations. As shown at the end
of this section, this inexact minimization suffices to claim convergence to a stationary
point of (P3). 

Similarly, in [S2] and [S3] the minimizations that give rise to $\bL[k+1]$ and $\bQ[k+1]$ 
are separable over their respective rows. For instance, the $l$-th row $\bl_l'$ of the
traffic subspace matrix $\bL:=[\bl_1,\ldots,\bl_L]'$ is updated as the solution of the
following ridge-regression problem
\begin{align}
\bl_l[k+1] = \arg\min_{\bl} \left[ \frac{1}{2} \|((\by_l^r)' -
\bl'\bQ'[k] - (\br^r_l)' \bA[k+1])\bm{\Omega}_l^r\|_2^2 + \frac{\lambda_{\ast}}{2} \|\bl\|_2^2
\right] \label{eq:ls_problem_bcd_l_l}
\end{align}
where $(\by_l^r)'$ and $(\br^r_l)'$ represent the $l$-th row of
$\bY$ and $\bR$, respectively. The $t$-th diagonal entry of the diagonal matrix
$\bm{\Omega}_l^r \in \mathbb{R}^{T \times T}$ 
is an indicator variable testing whether measurement $y_{l,t}$ is available. 
A similar regularized LS problem yields $\bq_t[k+1]$, $t=1,\ldots,T$; see Algorithm~\ref{tab:table_1}
for the details and a description of the overall BCD solver.
The novel batch scheme for unveiling network anomalies is 
less complex computationally than the accelerated
proximal gradient algorithm in~\cite{tit_exactrecovery_2012}, since Algorithm
\ref{tab:table_1}'s iterations are devoid of SVD computations.

\begin{algorithm}[t]
\caption{: Batch BCD algorithm for unveiling network anomalies} \small{
\begin{algorithmic}
	\STATE \textbf{input} $\cP_{\Omega}(\bY), \Omega, \bR, \lambda_{*},$ and $\lambda_1$.
    \STATE \textbf{initialize}
$\bL[1]$ and $\bQ[1]$ at random.
    \FOR {$k=1,2$,$\ldots$}
        \STATE {\bf [S1]}  \textbf{Update the anomaly map:}
        \FOR {$f=1,\ldots,F$}
        	  \STATE $\tilde{\by}_t^{(-f)}[k+1]=\mathbf{\Omega}_t(\by_t-\bL[k]\bq_t[k]-
        	  \sum_{f'=1}^{f-1}\br_{f'} a_{f',t}[k+1]-\sum_{f'=f+1}^{F}\br_{f'} a_{f',t}[k]), \quad t=1,\ldots,T$.
              \STATE $a_{f,t}[k+1]=\textrm{sign}(\br_{f}'\tilde{\by}_t^{(-f)}[k+1])
              \big[|\br_{f}'\tilde{\by}_t^{(-f)}[k+1]|-\lambda_1\big]_+/\|\br_{f}\|_2, \quad t=1,\ldots,T$.
        \ENDFOR
        \STATE $\bA[k+1] = \big[[a_{1,1}[k+1],\ldots,a_{F,1}[k+1]]', \ldots, 
[a_{1,T}[k+1],\ldots,a_{F,T}[k+1]]'\big]$.

        \STATE {\bf [S2]}  \textbf{Update the nominal traffic subspace:}
        \STATE $\bl_l[k+1]= \left(\lambda_{\ast} \bI_{\rho} + \bQ'[k]
        \mathbf{\Omega}_{l}^r \bQ[k] \right)^{-1} \bQ'[k]
        \mathbf{\Omega}_{l}^r
        (\by_l^r - \bA'[k+1]\br_l^r) , \quad l=1,\ldots,L$.

        \STATE $\bL[k+1] = [\bl_1[k+1], \ldots, \bl_L[k+1]]'$.

        \STATE {\bf [S3]}  \textbf{Update the projection coefficients:}
        \STATE $\bq_t[k+1] = \left( \lambda_{\ast} \bI_{\rho} + \bL'[k+1]
        \mathbf{\Omega}_t \bL[k+1] \right)^{-1} \bL'[k+1] \mathbf{\Omega}_t(\by_t
        - \bR \ba_t[k+1]), \quad t=1,\ldots,T$.
        \STATE $\bQ[k+1] = [\bq_1[k+1], \ldots, \bq_T[k+1]]'$.

    \ENDFOR
    \RETURN $\hat{\bA}:=\bA[\infty]$ and $\hat{\bX}:=\bL[\infty]\bQ'[\infty]$.
\end{algorithmic}}
\label{tab:table_1}
\end{algorithm}

Despite being non-convex and non-differentiable, (P3) has favorable structure which
facilitates convergence of the iterates generated by Algorithm \ref{tab:table_1}. 
Specifically, the resulting cost is convex in
each block variable when the rest are fixed. 
The non-smooth $\ell_1$-norm is also separable over the entries of its matrix argument. Accordingly, 
\cite[Theorem 5.1]{tseng_cnvg_bcd} guarantees convergence of the BCD algorithm to a stationary point of (P3).
This result together with Proposition~\ref{prop:prop_1} establishes the next claim.

\begin{proposition}\label{prop:prop_2}
If a subsequence $\{\bX[k]:=\bL[k]\bQ'[k],\bA[k]\}$ of iterates generated by
Algorithm~\ref{tab:table_1} satisfies
$\|\cP_{\Omega}(\bY-\bX[k]-\bR\bA[k])\| \leq \lambda_*$, then it
converges to the optimal solution set of (P1) as $k\rightarrow \infty$.
\end{proposition}

In practice, it is desirable to monitor anomalies in real time
and accomodate time-varying traffic routes.
These reasons motivate devising algorithms for \textit{dynamic} anomalography, 
the subject dealt with next.


\section{Dynamic Anomalography}\label{sec:online_alg}

Monitoring of large-scale IP networks necessitates collecting massive amounts 
of data which far outweigh the ability of modern computers to store and analyze them 
in real time. In addition, nonstationarities due to routing changes and missing data
further challenge identification of anomalies. In dynamic networks routing tables
are constantly readjusted to effect traffic load balancing and avoid
congestion caused by e.g., traffic anomalies or network infrastructure failures. 
To account for slowly time-varing routing tables, let $\bR_t \in
\mathbb{R}^{L \times F}$ denote the routing matrix at time $t$\footnote{Fixed
size routing matrices $\bR_t$ are considered here for convenience, where $L$ and
$F$ correspond to upper bounds on the number of physical links and flows transported
by the network, respectively. If at time $t$ some links are not used at all, or, less 
than $F$ flows are present, the corresponding rows
and columns of $\bR_t$ will be identically zero.}. In this dynamic setting, the partially observed 
link counts at time $t$ adhere to [cf. \eqref{eq:Y_modf}]
\begin{equation}
\cP_{\Omega_t}(\by_t)=\cP_{\Omega_t}(\bx_t + \bR_t\ba_t +
\bv_t),~t=1,2,\ldots \label{eq:y_t_dynamic}
\end{equation}
where the link-level traffic $\bx_t:=\bR_t\bz_t$, for $\bz_t$ from the 
(low-dimensional) traffic subspace. In general, routing
changes may alter a link load considerably by e.g.,
routing traffic completely away from a specific link. Therefore, even
though the network-level traffic vectors $\{\bz_t\}$ live in a low-dimensional
subspace, the same may not be true for the link-level traffic $\{\bx_t\}$
when the routing updates are major and frequent.
In backbone networks however, routing changes are sporadic relative
to the time-scale of data acquisition used for network monitoring tasks. 
For instance, data collected from the operation of Internet-2 
network  reveals that only a few rows of 
$\bR_t$ change per week~\cite{Internet2}. It is thus safe to assume that
$\{\bx_t\}$ still lies in a low-dimensional subspace, and exploit the temporal correlations 
of the observations to identify the anomalies.

On top of the previous arguments, in practice link measurements are acquired sequentially in
time, which motivates updating previously obtained estimates rather than
re-computing new ones from scratch each time a new datum becomes
available. The goal is then to recursively estimate $\{\hat{\bx}_t,\hat{\ba}_t\}$ at time
$t$ from historical observations $\{\cP_{\Omega_\tau}(\by_\tau),\Omega_\tau\}_{\tau=1}^t$,
naturally placing more importance to recent measurements. To this end, one
possible adaptive counterpart to~(P3) is the exponentially-weighted 
LS estimator found by minimizing the empirical cost
\begin{align}
\min_{\{\bL,\bQ,\bA\}} \sum_{\tau=1}^t \beta^{t-\tau}\left[ \frac{1}{2}
\|\cP_{\Omega_\tau}(\by_\tau-\bL\bq_\tau-\bR_\tau\ba_\tau) \|_2^2 + \frac{\lambda_{\ast}}{2
\sum_{u=1}^t \beta^{t-u}} \|\bL\|_F^2 +  \frac{\lambda_{\ast}}{2} \|\bq_\tau\|_2^2
+ \lambda_1 \|\ba_\tau\|_1 \right] \label{eq:adaptive_v1_est_lq}
\end{align}
in which $ 0< \beta \leq 1$ is the so-termed forgetting factor.  When $\beta<1$ 
data in the distant past are 
exponentially downweighted,  which facilitates tracking network 
anomalies in nonstationary environments. In the case of static
routing ($\bR_t=\bR, t=1,2,\ldots$) and infinite memory $(\beta=1)$, 
the formulation~\eqref{eq:adaptive_v1_est_lq} coincides
with the batch estimator (P3). This is the reason for the time-varying
factor weighting $\|\bL\|_F^2$.


\subsection{Tracking network anomalies}\label{subsec:alternating_minimization}

Towards deriving a real-time, computationally efficient, and recursive
solver of~\eqref{eq:adaptive_v1_est_lq}, an alternating
minimization method is adopted in which iteration $k$ coincides with the time
scale $t$ of data acquisition. A justification in terms of minimizing
a suitable approximate cost function is discussed in detail in Section \ref{subsec:cnvg_analysis}.
Per time instant $t$, a new datum
$\{\cP_{\Omega_t}(\by_t),\Omega_t\}$ is drawn and $\{\bq_t,\ba_t\}$ are jointly
estimated via
\begin{align}
\{\bq[t],\ba[t]\} = \arg\min_{\{\bq,\ba\}} \left[ \frac{1}{2}
\|\cP_{\Omega_t}(\by_t-\bL[t-1]\bq - \bR_t\ba)\|_2^2 +
\frac{\lambda_{\ast}}{2}
\|\bq\|_2^2 + \lambda_1 \|\ba\|_1 \right]. \label{eq:rec_est_q}
\end{align}
It turns out that~\eqref{eq:rec_est_q} can be efficiently solved. Fixing $\ba$
to carry out the minimization with respect to $\bq$ first, one is left with an 
$\ell_2$-norm regularized LS (ridge-regression) problem
\begin{align}
\bq[t] = &\arg\min_{\bq} \left[ \frac{1}{2} \|\cP_{\Omega_t}(\by_t -
\bL[t-1]\bq - \bR_t\ba)\|_2^2 + \frac{\lambda_{\ast}}{2} \|\bq\|^2 \right]\nonumber\\
&=\left( \lambda_{\ast} \bI_{\rho} +  \bL'[t-1] \mathbf{\Omega}_t \bL[t-1]
\right)^{-1}
\bL'[t-1]\cP_{\Omega_t}(\by_t - \bR_t\ba).  \label{eq:q_ls}
\end{align}
Note that $\bq[t]$ is an affine function of $\ba$, and the update rule for $\bq[t]$
is not well defined until $\ba$ is replaced with $\ba[t]$. Towards
obtaining an expression for $\ba[t]$, define $\bD[t]:=\left(\lambda_{\ast} 
\bI_{\rho}+\bL[t-1]\mathbf{\Omega_t} \bL'[t-1]\right)^{-1} \bL'[t-1]$ for notational
convenience, and substitute~\eqref{eq:q_ls} back into~\eqref{eq:rec_est_q} to arrive at the Lasso estimator
\begin{align}
\ba[t] = &\arg\min_{\ba} \left[ \frac{1}{2} \|\bF[t](\by_t -
\bR_t\ba)\|_2^2 + \lambda_1 \|\ba\|_1 \right] \label{eq:a_lasso}
\end{align}
where $\bF[t]:= \left[\mathbf{\Omega}_t - \mathbf{\Omega}_t\bL[t-1] \bD[t] 
\mathbf{\Omega}_t, \sqrt{\lambda_{\ast}} \mathbf{\Omega}_t \bD'[t]\right]'$.
The diagonal matrix $\bm{\Omega}_t$ was defined in Section \ref{ssec:bcd_alg},
see the discussion after \eqref{eq:lasso}.

In the second step of the alternating-minimization
scheme, the updated subspace matrix $\bL[t]$ is obtained
by minimizing \eqref{eq:adaptive_v1_est_lq} with respect to $\bL$, 
while the optimization variables $\{\bq_{\tau},\ba_{\tau}\}_{\tau=1}^t$ are fixed and take the values 
$\{\bq[\tau],\ba[\tau]\}_{\tau=1}^t$. This yields
\begin{align}
\bL[t] = \arg\min_{\bL} \left[\frac{\lambda_{\ast}}{2} \|\bL\|_F^2+ \sum_{\tau=1}^t \beta^{t-\tau}
\frac{1}{2} \|\cP_{\Omega_\tau}(\by_\tau -
\bL\bq[\tau] - \bR_\tau\ba[\tau])\|_2^2\right]. \label{eq:rec_est_L}
\end{align}
Similar to the batch case, \eqref{eq:rec_est_L} decouples over the rows of $\bL$ which are
obtained in parallel via 
\begin{align}
\bl_{l}[t] = &\arg\min_{\bl} \left[ \frac{\lambda_{\ast}}{2} \|\bl\|^2+\sum_{\tau=1}^t \beta^{t-\tau}
\omega_{l,\tau}(y_{l,\tau} - \bl'\bq[\tau] - (\br_{l,\tau}^r)'\ba[\tau])^2\right],\quad l=1,\ldots,L 
\label{eq:rec_est_li}
\end{align}
where $\omega_{l,\tau}$ denotes the $l$-th diagonal entry of $\bm{\Omega}_{\tau}$.
For $\beta=1$, subproblems \eqref{eq:rec_est_li} can be efficiently
solved using the RLS algorithm~\cite{Solo_Adaptive_Book}. Upon
defining $\bs_l[t]:=\sum_{\tau=1}^t\beta^{t-\tau}\omega_{l,\tau}
(y_{l,\tau} - \br'_{l,\tau}\ba[\tau])\bq[\tau]$, 
$\bH_l[t]:=\sum_{\tau=1}^t\beta^{t-\tau}\omega_{l,\tau}\bq[\tau]\bq'[\tau]+
\lambda_\ast\bI_\rho$, and $\bM_{l}[t]:=\bH_l^{-1}[t]$, with $\beta=1$ one simply updates 
\begin{align}
\bs_{l}[t]{}={} & \bs_{l}[t-1] +\omega_{l,t}(y_{l,t}-\br'_{l,t}\ba[t]) \bq[t]\nonumber\\
\bM_{l}[t] {}={} & \bM_{l}[t-1] 
-\omega_{l,t}\frac{\bM_l[t-1]\bq[t]\bq'[t]\bM_l[t-1]}{1+\bq'[t]\bM_l[t-1]\bq[t]}\nonumber
\end{align}
and forms $\bl_{l}[t]=\bM_{l}[t]\bs_{l}[t]$, for $l=1,\ldots,L$.

However, for $0< \beta<1$  the
regularization term $(\lambda_{\ast}/2) \|\bl\|^2$ in \eqref{eq:rec_est_li} makes
it impossible to express $\bH_l[t]$
in terms of $\bH_l[t-1]$ plus a rank-one correction. Hence, one cannot resort to the
matrix inversion lemma and update $\bM_{l}[t]$ with quadratic complexity only.
Based on direct inversion of $\bH_l[t]$, $l=1,\ldots,L$, the overall recursive 
algorithm for tracking network anomalies is tabulated under Algorithm~\ref{tab:table_2}. 
The per iteration cost of the $L$ inversions (each $\mathcal{O}(\rho^3)$, which could be further reduced
if one leverages also the symmetry of $\bH_l[t]$) is affordable for moderate number of links, because $\rho$
is small when estimating low-rank traffic matrices. Still, for those settings 
where computational complexity reductions are at a premium,
an online stochastic gradient descent algorithm is described in Section \ref{ssec:nesterov}.

\begin{algorithm}[t]
\caption{: Online algorithm for tracking network anomalies} \small{
\begin{algorithmic}
	\STATE \textbf{input} $ 
	\{\cP_{\Omega_t}(\by_t),\mathbf{\Omega}_t,\bR_t\}_{t=1}^{\infty} ,\beta, 
	\lambda_{\ast},$ and $\lambda_1$.
    \STATE \textbf{initialize} $\bG_{l}[0]=\mathbf{0}_{\rho\times \rho}$,
    $\bs_{l}[0]=\mathbf{0}_{\rho},~l=1,...,L$, and $\bL[0]$ at
    random.
    \FOR {$t=1,2$,$\ldots$}
                \STATE $\bD[t] = \left(\lambda_{\ast} \bI_{\rho} + \bL'[t-1]
                \mathbf{\Omega}_t \bL[t-1]\right)^{-1} \bL'[t-1]$.
                \STATE $\bF[t]= \left[\mathbf{\Omega}_t -
                \mathbf{\Omega}_t\bL[t-1]\bD[t]\mathbf{\Omega}_t, \sqrt{\lambda_{\ast}}
                \mathbf{\Omega}_t \bD'[t]\right]'$.
                \STATE $\ba[t] = \arg\min_{\ba} \left[\frac{1}{2}
                \|\bF[t](\by_t- \bR_t\ba)\|^2 + \lambda_1 \|\ba\|_1 \right]$.
                \STATE $\bq[t] = \bD[t] \mathbf{\Omega}_t(\by_t-\bR_t \ba[t])$.
                \STATE $\bG_{l}[t] = \beta \bG_{l}[t-1] + \omega_{l,t}
                \bq[t]\bq[t]', \quad l=1,\ldots,L$.
                \STATE $\bs_{l}[t]=\beta \bs_{l}[t-1] +
                \omega_{l,t}(y_{l,t}-\br'_{l,t}\ba[t]) \bq[t], \quad
                l=1,\ldots,L$.
                 \STATE $\bl_{l}[t] = \left(\bG_{l}[t] + \lambda_{\ast}
                 \bI_{\rho}\right)^{-1} \bs_{l}[t], \quad l=1,...,L$.
                \RETURN  $\hat{\ba}_t:=\ba[t]$ and $\hat{\bx}_t := \bL[t] \bq[t]$.
    \ENDFOR
\end{algorithmic}}
\label{tab:table_2}
\end{algorithm}

\begin{remark}[Robust subspace trackers]
Algorithm \ref{tab:table_2} is closely related to timely robust
subspace trackers, which aim at estimating a low-rank subspace $\bL$
from grossly corrupted and possibly incomplete data, namely $\cP_{\Omega_t}(\by_t)=\cP_{\Omega_t}
(\bL\bq_t + \ba_t +\bv_t),~t=1,2,\ldots$. In the absence of sparse `outliers' $\{\ba_t\}_{t=1}^\infty$,
an online algorithm based on incremental gradient descent on the Grassmannian manifold of subspaces
was put forth in~\cite{onlinetracking_bolzano10}. The second-order 
RLS-type algorithm in~\cite{petrels_chi12} extends the seminal projection
approximation subspace tracking algorithm~\cite{yang95} to handle missing data.
When outliers are present, robust counterparts can be found 
in~\cite{Vaswani_Allerton_11,increm_grad_grasmannian_bolzano12,gonzalo_rpca}.
Relative to all aforementioned works, the estimation problem here is more challenging due to the presence
of the fat (compression) matrix $\bR_t$; see~\cite{tit_exactrecovery_2012} for fundamental
identifiability issues related to the model~\eqref{eq:Y_modf}.
\end{remark}


\subsection{Convergence Analysis}
\label{subsec:cnvg_analysis}

This section studies the convergence of the iterates generated by
Algorithm~\ref{tab:table_2}, for the infinite memory special case i.e., when $\beta=1$. 
Upon defining the function
\begin{align}
g_t(\bL,\bq,\ba):=  \frac{1}{2}
\|\cP_{\Omega_t}(\by_t-\bL\bq - \bR_t\ba)\|_2^2 +
\frac{\lambda_{\ast}}{2}
\|\bq\|_2^2 + \lambda_1 \|\ba\|_1 \nonumber 
\end{align}
in addition to $\ell_t(\bL):=\min_{\{\bq,\ba\}}g_t(\bL,\bq,\ba) $,
the online solver of Section \ref{subsec:alternating_minimization}
aims at minimizing the following \textit{average} cost function  at time $t$
\begin{align}
C_t(\bL) := \frac{1}{t} \sum_{\tau=1}^t \ell_\tau(\bL) + \frac{\lambda_{\ast}}{2t}
\|\bL\|_F^2. \label{eq:cost_target}
\end{align}
Normalization (by $t$) ensures that the cost function does not grow unbounded as
time evolves. For any finite $t$, \eqref{eq:cost_target} it is essentially identical to the batch
estimator in (P3) up to a scaling, which does not affect the value of
the minimizers. Note that as time evolves, minimization of $C_t$
becomes increasingly complex computationally. Even evaluating
$C_t$ is challenging for large $t$, since it entails solving $t$ Lasso problems 
to minimize all $g_\tau$ and define 
the functions $\ell_\tau$, $\tau=1,\ldots,T$. Hence, at time $t$ the subspace estimate $\bL[t]$ is
obtained by minimizing the \textit{approximate} cost function
\begin{align}
\hat{C}_t(\bL) = \frac{1}{t} \sum_{\tau=1}^t g_\tau(\bL,\bq[\tau],\ba[\tau]) +
\frac{\lambda_{\ast}}{2t} \|\bL\|_F^2 \label{eq:cost_apx}
\end{align}
in which $\{\bq[t],\ba[t]\}$ are obtained based on the prior
subspace estimate $\bL[t-1]$ after solving [cf. \eqref{eq:rec_est_q}]
\begin{align}
\{\bq[t],\ba[t]\}=\arg\min_{\{\bq,\ba\}} g_t(\bL[t-1],\bq,\ba).
\label{eq:a&q_givenL_t-1}
\end{align}
Obtaining $\bq[t]$ this way resembles the
projection approximation adopted in~\cite{yang95}, and can only be evaluated after $\ba[t]$ is obtained 
[cf. \eqref{eq:q_ls}]. Since $\hat{C}_t(\bL)$ is a smooth convex function, the minimizer 
$\bL[t]=\arg\min_{\bL}\hat{C}_t(\bL)$ is the solution of the quadratic equation $\nabla \hat{C}_t(\bL[t]) 
=\mathbf{0}_{L\times \rho}$. 

So far, it is apparent that the approximate cost function $\hat{C}_t(\bL[t])$ 
overestimates the target cost $C_t(\bL[t])$, for $t=1,2,\ldots$. 
However, it is not clear whether the dictionary
iterates $\{\bL[t]\}_{t=1}^\infty$ converge, and most importantly, how well can they optimize the target
cost function $C_t$. The good news is that $\hat{C}_t(\bL[t])$ 
asymptotically approaches $C_t(\bL[t])$, and the subspace iterates null
$\nabla C_t(\bL[t])$ as well, both as $t\to\infty$. 
The latter result is summarized in the next proposition, which is proved in the next section.

\begin{proposition}\label{th:thorem_1}
Assume that: a1) $\{\Omega_t\}_{t=1}^\infty$ and $\{\by_t\}_{t=1}^\infty$
are independent and identically distributed (i.i.d.) random processes; 
a2) $\|\cP_{\Omega_t}(\by_t)\|_{\infty}$ is uniformly
bounded; a3) iterates $\{\bL[t]\}_{t=1}^\infty$ are in a compact set $\mathcal{L} \subset
\mathbb{R}^{L \times \rho}$; a4) $\hat{C}_t(\bL)$ is positive
definite, namely $\lambda_{\min}\left[\nabla^2 \hat{C}_t(\bL) \right] \geq c
$ for some $c>0$; and a5) the Lasso~\eqref{eq:a_lasso} has a unique solution. 
Then $\lim_{t \rightarrow \infty} \nabla
C_t(\bL[t]) = \mathbf{0}_{L\times\rho}$ almost surely (a.s.), i.e., the subspace iterates $\{\bL[t]\}_{t=1}^\infty$
asymptotically coincide with the stationary points of the batch problem (P3).
\end{proposition}

To clearly delineate the scope of the analysis, 
it is worth commenting on the assumptions a1)-a5) and the 
factors that influence their satisfaction. Regarding a1), the acquired
data is assumed statistically independent across time as it is customary
when studying the stability and performance of online (adaptive) algorithms~\cite{Solo_Adaptive_Book}.
Still, in accordance with the adaptive filtering folklore, as $\beta\to 1$
the upshot of the analysis based on i.i.d. data extends accurately to the pragmatic
setting whereby the link-counts and missing data patterns exhibit
spatiotemporal correlations. Uniform boundedness of $\cP_{\Omega_t}(\by_t)$ [cf. a2)] is
satisfied in practice, since the traffic is always limited by the (finite) 
capacity of the physical links. The bounded subspace requirement in a3)
is a technical assumption that simplifies the arguments of the ensuing proof,
and has been corroborated via computer simulations. It is apparent that 
the sampling set $\Omega_t$ plays a key
role towards ensuring that a4) and a5) are satisfied. Intuitively, if the missing entries
tend to be only few and somehow uniformly distributed across links and time, they
will not markedly increase coherence of the regression matrices
$\bF[t]\bR_t$, and thus compromise the uniqueness of the Lasso solutions. This
also increases the likelihood that $\nabla^2 \hat{C}_t(\bL)= \frac{\lambda_{\ast}}{t}
\bI_{L\rho} +\frac{1}{t}\sum_{\tau=1}^t (\bq[\tau] \bq'[\tau]) \otimes \mathbf{\Omega}_\tau \succeq
c\bI_{L\rho}$ holds. As argued in~\cite{mairalonlinelearning},
if needed one could incorporate additional regularization terms in the cost function 
to enforce a4) and a5). Before moving on to the
proof, a remark is in order.

\begin{remark}[Performance guarantees] In line with Proposition~\ref{prop:prop_2}, 
one may be prompted to ponder whether the
online estimator offers the performance guarantees of the nuclear-norm regularized
estimator (P1), for which stable/exact recovery have been well documented e.g.,
in~\cite{tit_exactrecovery_2012,zlwcm10,CLMW09}. Specifically, given the learned
traffic subspace $\bar{\bL}$ and the corresponding $\bar{\bQ}$ and $\bar{\bA}$
[obtained via~\eqref{eq:rec_est_q}] over a time window of size $T$, is
$\{\hat{\bX}:=\bar{\bL}\bar{\bQ}',\hat{\bA}:=\bar{\bA}\}$ an optimal solution
of (P1) when $T \rightarrow \infty$? This in turn requires asymptotic analysis
of the optimality conditions for (P1), and is left for future research.
Nevertheless, empirically the online estimator attains the
performance of (P1), as evidenced by the numerical tests in Section~\ref{sec:sims}.
\end{remark}


\subsection{Proof of Proposition~\ref{th:thorem_1}}
\label{subsec:proof}
The main steps of the proof are inspired by~\cite{mairalonlinelearning}, which
studies convergence of an online dictionary learning algorithm using 
the theory of martingale sequences; see e.g.,~\cite{Ljung}. 
However, relative to~\cite{mairalonlinelearning} the problem
here introduces several distinct elements including: i) missing data with a
time-varying pattern $\Omega_t$; ii) a non-convex bilinear term where the 
tall subspace matrix $\bL$ plays a role similar
to the fat dictionary in~\cite{mairalonlinelearning}, but the multiplicative
projection coefficients here are not sparse; and iii) the additional
bilinear terms $\bR_t\ba_t$ which entail sparse coding of $\ba_t$ as in~\cite{mairalonlinelearning},
but with a known regression (routing) matrix.
Hence, convergence analysis becomes more challenging and demands, in part, for a new treatment. 
Accordingly, in the sequel emphasis will be placed on the novel aspects specific to the problem
at hand.

The basic structure of the proof consists of three preliminary lemmata, which
are subsequently used to establish that $\lim_{t \rightarrow \infty} \nabla
C_t(\bL[t]) = \mathbf{0}_{L\times \rho}$ a.s. through a simple argument. 
The first lemma deals with regularity properties of
functions $\hat{C}_t$ and $C_t$, which will come handy later on; see Appendix
A for a proof.

\begin{lemma}\label{lem:lemma_2}
If a2) and a5) hold, then the functions: i)
$\{\ba_t(\bL),\bq_t(\bL)\}=\arg\min_{\{\bq,\ba\}} g_t(\bL,\bq,\ba)$, ii)
$g_t(\bL,\bq[t],\ba[t])$,
iii) $\ell_t(\bL)$, and iv) $\nabla\ell_t(\bL)$ are Lipschitz continuous for $\bL \in
\mathcal{L}$ ($\mathcal{L}$ is a compact set), with constants independent of $t$.
\end{lemma}

\noindent The next lemma (proved in Appendix B) asserts that the distance between two subsequent
traffic subspace estimates vanishes as $t\to\infty$, a property that
will be instrumental later on when establishing that $\hat{C}_t(\bL[t])-C_t(\bL[t])\to 0$
a.s.

\begin{lemma}\label{lem:lemma_3}
If a2)-a5) hold, then $\|\bL[t+1]-\bL[t]\|_F = \mathcal{O}(1/t)$.
\end{lemma}

\noindent The previous lemma by no means implies that the subspace iterates converge, which is
a much more ambitious objective that may not even hold under the current assumptions. 
The final lemma however, asserts that the cost sequence indeed converges with probability
one; see Appendix C for a proof. 

\begin{lemma}\label{lem:lemma_4}
If a1)-a5) hold, then $\hat{C}_t(\bL[t])$ converges a.s. Moreover,
$\hat{C}_t(\bL[t])-C_t(\bL[t])\to 0$ a.s.
\end{lemma}

Putting the pieces together, in the sequel it is shown that the sequence
$\{\nabla \hat{C}_t(\bL[t])-\nabla C_t(\bL[t])\}_{t=1}^\infty$ converges a.s. to
zero, and since $\nabla \hat{C}_t(\bL[t])=\mathbf{0}_{L\times \rho}$ by algorithmic
construction, the subspace iterates $\{\bL[t]\}_{t=1}^\infty$ coincide 
with the stationary points of the target cost function $C_t$.
To this end, it suffices to prove that every convergent \textit{subsequence} nulls the
gradient $\nabla C_t$ asymptotically, which in turn implies that the entire sequence converges 
to the set of stationary points of the batch problem (P3).

Since $\mathcal{L}$ is compact by virtue of a3), one can always pick a convergent subsequence
$\{\bL[t]\}_{t=1}^\infty$ whose limit point is $\bL^*$, say\footnote{Formally, the subsequence
should be denoted as $\{\bL[t(i)]\}_{i=1}^\infty$, but a slight
abuse of notation is allowed for simplicity.}. Consider the
positive-valued decreasing sequence $\{\alpha_t\}_{t=1}^\infty$ 
(that necessarily converges to zero), and recall
that $\hat{C}_t(\bL[t]+\alpha_t \bU) \geq C_t(\bL[t]+\alpha_t \bU)$ for any $\bU \in
\mathbb{R}^{L \times \rho}$. From the mean-value theorem and for arbitrary $\bU$, expanding both
sides of the inequality around the point $\bL[t]$ one arrives at
\begin{align}
&\hat{C}_t(\bL[t]) + \alpha_t \tr\{ \bU' \nabla \hat{C}_t(\bL[t])\} +
\frac{1}{2}
\alpha_t^2
\tr\{\bU' \nabla^2 \hat{C}_t(\mathbf{\Theta}_1) \bU \} \geq \nonumber  \\
&\hspace{55mm}C_t(\bL[t]) + \alpha_t \tr\{\bU'
\nabla C_t(\bL[t])\} + \frac{1}{2} \alpha_t^2 \tr\{\bU' \nabla^2
C_t(\mathbf{\Theta}_2) \bU
\}\nonumber
\end{align}
for some $\mathbf{\Theta}_1,\mathbf{\Theta}_2 \in \mathbb{R}^{L \times \rho}$
and all $t$. Taking limit as $t\to\infty$ and applying Lemma~\ref{lem:lemma_4}
it follows that
\begin{align}
& \lim_{t \rightarrow \infty}\tr\{ \bU' (\nabla \hat{C}_t(\bL[t]) - \nabla
C_t(\bL[t]))\} +
\lim_{t \rightarrow \infty}\frac{1}{2}
\alpha_t
\tr\{\bU' (\nabla^2 \hat{C}_t(\mathbf{\Theta}_1) - \nabla^2
C_t(\mathbf{\Theta}_2)) \bU \} \geq 0. \label{eq:ineq_taylor_limit}
\end{align}
One can readily show that $\nabla^2 \hat{C}_t(\mathbf{\Theta}_1)=\frac{1}{t}
\sum_{\tau=1}^t (\bq[\tau] \bq'[\tau]) \otimes \mathbf{\Omega}_\tau +
\frac{\lambda_{\ast}}{t}\bI_{L\rho}$ is bounded since $\bL[t]$ is uniformly
bounded [cf. a2)]. Consequently, $\lim_{t \rightarrow \infty}\frac{1}{2}
\alpha_t \tr\{\bU' (\nabla^2 \hat{C}_t(\mathbf{\Theta}_1)) \bU \} =0$. Furthermore,
since $\nabla \ell_\tau$ is Lipschitz as per Lemma~\ref{lem:lemma_2},
$\nabla C_t$ is Lipschitz as well
and it follows that $\lim_{t \rightarrow \infty}\frac{1}{2}
\alpha_t\tr\{\bU' \nabla^2
C_t(\mathbf{\Theta}_2) \bU \} = 0$. All in all, the second term
in~\eqref{eq:ineq_taylor_limit} vanishes and one is left with
\begin{align}
& \lim_{t \rightarrow \infty}\tr\{ \bU' (\nabla \hat{C}_t(\bL_t) - \nabla
C_t(\bL_t))\}  \geq 0. \label{eq:ineq_taylor_limit_grad}
\end{align}
Because $\bU \in \mathbb{R}^{L \times \rho}$ is arbitrary,~\eqref{eq:ineq_taylor_limit_grad} 
can only hold if $\lim_{t
\rightarrow \infty} (\nabla \hat{C}_t(\bL[t]) - \nabla
C_t(\bL[t])) = \mathbf{0}_{L \times \rho}$ a.s., which completes the proof.\hfill$\blacksquare$


\section{Further Algorithmic Issues}\label{sec:alg_issues}

For completeness, this section outlines a couple of additional algorithmic 
aspects relevant to anomaly detection
in \textit{large-scale} networks. Firstly, a lightweight first-order algorithm is developed 
as an alternative to Algorithm \ref{tab:table_2}, which relies on fast
Nesterov-type gradient updates for the traffic subspace. 
Secondly, the possibility of developing distributed algorithms for dynamic anomalography
is discussed.


\subsection{Fast stochastic-gradient algorithm}
\label{ssec:nesterov}

Reduction of the computational complexity in updating the traffic subspace $\bL$
is the subject of this section. The basic alternating minimization
framework in Section \ref{subsec:alternating_minimization} will be retained,
and the updates for $\{\bq[t],\ba[t]\}$ will be identical to those tabulated under
Algorithm \ref{tab:table_2}. However, instead of solving an
unconstrained quadratic program per iteration to obtain $\bL[t]$ [cf. \eqref{eq:rec_est_L}],
the refinements to the subspace estimate will be given by a (stochastic)
gradient algorithm. 

As discussed in Section \ref{subsec:cnvg_analysis}, 
in Algorithm \ref{tab:table_2} the subspace estimate $\bL[t]$
is obtained by minimizing the empirical cost function 
$\hat{C}_t(\bL)=(1/t)\sum_{\tau=1}^t f_\tau(\bL)$, where
\begin{equation}\label{eq:f_t}
f_t(\bL):=\frac{1}{2}\|\bm{\Omega}_t(\bby_t-\bL\bbq[t]-\bR_t\bba[t])\|_2^2+
\frac{\lambda_*}{2t}\|\bL\|_F^2 +
\frac{\lambda_{\ast}}{2}
\|\bq[t]\|_2^2 + \lambda_1 \|\ba[t]\|_1, \quad t=1,2,\ldots
\end{equation}
By the law of large numbers, if data 
$\{\cP_{\Omega_t}(\by_t)\}_{t=1}^{\infty}$ are stationary, 
solving $\min_{\bL}\lim_{t\to\infty}\hat{C}_t(\bL)$  
yields the desired minimizer of the \textit{expected} cost $\mathbb{E}[C_t(\bL)]$, 
where the expectation is taken with respect to the unknown probability 
distribution of the data. A standard approach to achieve this same goal -- typically with
reduced computational complexity -- is to drop the expectation (or the sample averaging
operator for that matter), and update the nominal traffic subspace via a stochastic gradient
iteration~\cite{Solo_Adaptive_Book}
\begin{align}\label{eq:stochastic_gradient}
\bL[t]{}={}&\arg\min_{\bL}Q_{(1/\tilde{\mu}[t]),t}(\bL,\bL[t-1])\nonumber\\
={}&\bL[t-1]-\tilde{\mu}[t]\nabla f_t(\bL[t-1])
\end{align}
where $\tilde{\mu}[t]$ is a stepsize, $Q_{\mu,t}(\bL_1,\bL_2):=f_t(\bL_2)+\langle\bL_1-\bL_2,\nabla
f_t(\bL_2)\rangle +\frac{\mu}{2}\|\bL_1-\bL_2\|_f^2$, and $\nabla
f_t(\bL)=-\bm{\Omega}_t(\bby_t-\bL\bbq[t]-\bR_t\bba[t])\bbq'[t]+(\lambda_*/t)\bL$.
In the context of adaptive filtering, stochastic gradient algorithms such as \eqref{eq:f_t} 
are known to converge typically slower than RLS. This is expected
since RLS can be shown to be an instance of 
Newton's (second-order) optimization method~\cite{Solo_Adaptive_Book}.

Building on the increasingly popular \textit{accelerated} gradient methods for (batch)
smooth optimization~\cite{nesterov83,fista}, the idea here is to speed-up the 
learning rate of the estimated traffic subspace \eqref{eq:stochastic_gradient}, 
without paying a penalty in terms of computational complexity per iteration.
The critical difference between standard gradient algorithms and the so-termed 
Nesterov's variant,
is that the accelerated updates take the form $\bL[t]=\tilde\bL[t]-\tilde{\mu}[t]\nabla f_t(\tilde\bL[t])$, 
which relies on a judicious linear combination $\tilde\bL[t-1]$ of the previous pair of iterates 
$\{\bL[t-1],\bL[t-2]\}$. Specifically, the choice
$\tilde\bL[t]=\bL[t-1]+\frac{k[t-1]-1}{k[t]}\left(\bL[t-1]-\bL[t-2]\right)$, where
$k[t]=\left[1+\sqrt{4k^2[t-1]+1}\right]/2$, has been shown to significantly
accelerate batch gradient algorithms resulting in convergence rate no worse than 
$\mathcal{O}(1/k^2)$; see e.g.,~\cite{fista} and references therein. Using this acceleration
technique in conjunction with a backtracking stepsize rule~\cite{Bers}, a fast
online stochastic gradient algorithm for unveiling network anomalies is tabulated
under Algorithm \ref{tab:table_4}. Different from Algorithm \ref{tab:table_2}, no
matrix inversions are involved in the update of the traffic subspace $\bL[t]$.
Clearly, a standard (non accelerated) stochastic gradient
descent algorithm with backtracking stepsize rule is subsumed as a special
case, when $k[t]=1$, $t=0,1,2,\ldots$

\begin{algorithm}[t]
\caption{: Online stochastic gradient algorithm for unveiling network
anomalies} \small{
\begin{algorithmic}
	\STATE \textbf{input} $\{\bby_t,\bR_t,\bm{\Omega}_t\}_{t=1}^\infty,$
	$\rho, \lambda_*, \lambda_1,\eta>1$.
    \STATE \textbf{initialize}
$\bL[0]$ at random, $\mu[0]>0$, $\tilde\bL[1]:=\bL[0]$, and $k[1]:=1$.
    \FOR {$t=1,2$,$\ldots$}
                \STATE $\bD[t] = \left(\lambda_{\ast} \bI_{\rho} + \bL'[t-1]
                \mathbf{\Omega}_t \bL[t-1]\right)^{-1} \bL'[t-1]$
                \STATE $\bF'[t]:= \left[\mathbf{\Omega}_t -
                \mathbf{\Omega}_t\bL[t-1]
                                \bD[t]
                                \mathbf{\Omega}_t, \sqrt{\lambda_{\ast}}
                                \mathbf{\Omega}_t \bD'[t]
                                \right]$
                \STATE $\ba[t] = \arg\min_{\ba} \left[\frac{1}{2}
                \|\bF[t](\by_t
                - \bR_t\ba)\|^2 + \lambda_1 \|\ba\|_1 \right]$
                \STATE $\bq[t] = \bD[t] \mathbf{\Omega}_t
                (\by_t-\bR_t \ba_t)$
        \STATE Find the smallest nonnegative integer $i[t]$ such that with
        $\bar{\mu}:=\eta^{i[t]}\mu[t-1]$
        $$f_t(\tilde\bL[t]-(1/\bar{\mu})\nabla f_t(\tilde\bL[t]))\leq
        Q_{\bar{\mu},t}(\tilde\bL[t]-(1/\bar{\mu})\nabla
        f_t(\tilde\bL[t]),\tilde\bL[t])$$
        holds, and set $\mu[t]=\eta^{i[t]}\mu[t-1].$
        \STATE $\bL[t]=\tilde\bL[t]-(1/\mu[t])\nabla f_t(\tilde\bL[t]).$
        \STATE $k[t+1]=\frac{1+\sqrt{1+4k^2[t]}}{2}.$
        \STATE
        $\tilde\bL[t+1]=\bL[t]+\left(\frac{k[t]-1}{k[t+1]}\right)(\bL[t]-\bL[t-
        1]).$
    \ENDFOR
    \RETURN $\hat{\bbx}[t]:=\bL[t]\bbq[t], \hat{\bba}[t]:=\bba[t]$.
\end{algorithmic}}
\label{tab:table_4}
\end{algorithm}

Convergence analysis of Algorithm \ref{tab:table_4} is beyond the scope of
this paper, and will only be corroborated using computer simulations in 
Section \ref{sec:sims}. It is worth pointing out that since a non-diminishing
stepsize is adopted, asymptotically the iterates generated by Algorithm \ref{tab:table_4} 
will hover inside a  ball centered at the minimizer of the expected cost, with radius proportional
to the noise variance.


\subsection{In-network anomaly trackers}
\label{ssec:distributed}
Implementing Algorithms \ref{tab:table_1}-\ref{tab:table_4} presumes that network nodes continuously
communicate their local link traffic measurements to a central monitoring station, 
which uses their aggregation in 
$\{\cP_{\Omega_t}(\by_t)\}_{t=1}^{\infty}$ to unveil network anomalies.
While for the most part this is the prevailing operational paradigm adopted
in current network technologies, it is fair to say 
there are limitations associated with this architecture.
For instance, collecting all this information centrally may lead to 
excessive protocol overhead, especially when the rate of data acquisition is high
at the routers. Moreover, minimizing the exchanges of raw measurements
may be desirable to reduce unavoidable communication errors that translate to missing data. 
Performing the optimization in a centralized fashion raises robustness concerns as well,
since the central monitoring station represents an isolated point of failure. 

These reasons motivate devising \emph{fully-distributed} iterative algorithms
for dynamic anomalography in large-scale networks, embedding the network
anomaly detection functionality to the routers. In a nutshell, per iteration 
nodes carry out simple computational tasks locally, 
relying on their own link count measurements (a few entries
of the network-wide vector $\bby_t$ corresponding to the router links). 
Subsequently, local estimates are refined after exchanging messages 
only with directly connected neighbors, which facilitates percolation
of local information to the whole network. The end goal is 
for network nodes to consent on a global map of network anomalies, 
and attain (or at least come close to) the estimation performance
of the centralized counterpart which has all data $\{\cP_{\Omega_t}(\by_t)\}_{t=1}^{\infty}$
available. 

Relying on the alternating-directions method of multipliers (AD-MoM)
as the basic tool to carry out distributed optimization, a general
framework for in-network sparsity-regularized rank minimization was
put forth in a companion paper~\cite{tsp_rankminimization_2012}. In the
context of network anomaly detection, results therein
are encouraging yet there is ample room for improvement and immediate venues
for future research open up. For instance, the distributed
algorithms of~\cite{tsp_rankminimization_2012} can only tackle the batch formulation (P3), 
so extensions to a dynamic network setting, e.g., building on the ideas here to
devise distributed anomaly trackers seems natural. To obtain desirable tradeoffs
in terms of computational complexity and speed of convergence, developing
and studying algorithms for distributed optimization
based on Nesterov's acceleration techniques emerges as an exciting 
and rather pristine research direction; see~\cite{moura_dnesterov} 
for early work dealing with separable batch optimization.


\section{Performance Tests}
\label{sec:sims}
Performance of the proposed batch and online estimators is assessed in this
section via computer simulations using both synthetic and real network data. 

\begin{figure}[t]
\centering
  \centerline{\epsfig{figure=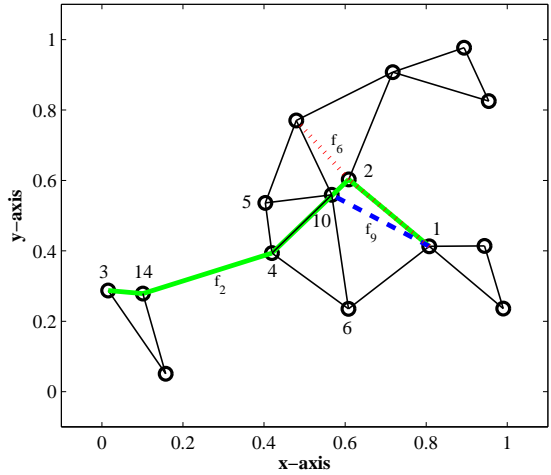,width=0.5\textwidth}
  }
\vspace{-5mm}\caption{Synthetic network topology graph, and the paths used for
routing three flows.}
  \label{fig:fig_nettolpology}
\end{figure}


\subsection{Synthetic network data tests}
\label{ssec:perf_synthetic}

\noindent\textbf{Synthetic network example}. A network of $N=15$ nodes is
considered as a realization of the random geometric graph model with
agents randomly placed on the unit square, and two agents link 
if their Euclidean distance is less than a prescribed communication
range of $d_c=0.35$; see Fig.~\ref{fig:fig_nettolpology}. The network graph is
bidirectional and comprises $L=52$ links, and $F=N(N-1)=210$ OD flows. For
each candidate OD pair, minimum hop count routing is considered to form the
routing matrix $\bR$. Entries of $\bv_t$ are i.i.d., 
zero-mean, Gaussian with variance $\sigma^2$; i.e.,
$v_{l,t}\sim \mathcal{N}(0,\sigma^2)$. Flow-traffic vectors $\bz_t$ are generated 
from the low-dimensional subspace $\bU\in\mathbb{R}^{F\times r}$ with i.i.d. entries 
$u_{f,i}\sim \mathcal{N}(0,1/F)$, and projection coefficients $w_{i,t}
\sim N(0,1)$ such that $\bz_t=\bU\bw_t$. Every entry of $\ba_t$ is randomly drawn
from the set $\{-1,0,1\}$, with ${\rm Pr} (a_{f,t}=-1)={\rm Pr}(a_{f,t}=1)=p/2$. 
Entries of $\bY$ are sampled uniformly at
random with probability $\pi$ to form the diagonal sampling matrix $\mathbf{\Omega}_t$. 
The observations at time instant $t$ are generated according to 
$\cP_{\Omega_t}(\by_t)=\mathbf{\Omega}_t (\bR\bz_t + 
\bR\ba_t + \bv_t)$.
Unless otherwise stated, $r=2$, $\rho=5$, and $\beta=0.99$ are used throughout. 
Different values of
$\sigma$, $p$ and $\pi$ are tested.

\begin{figure}[t]
\centering
\begin{tabular}{cc}
     \epsfig{file=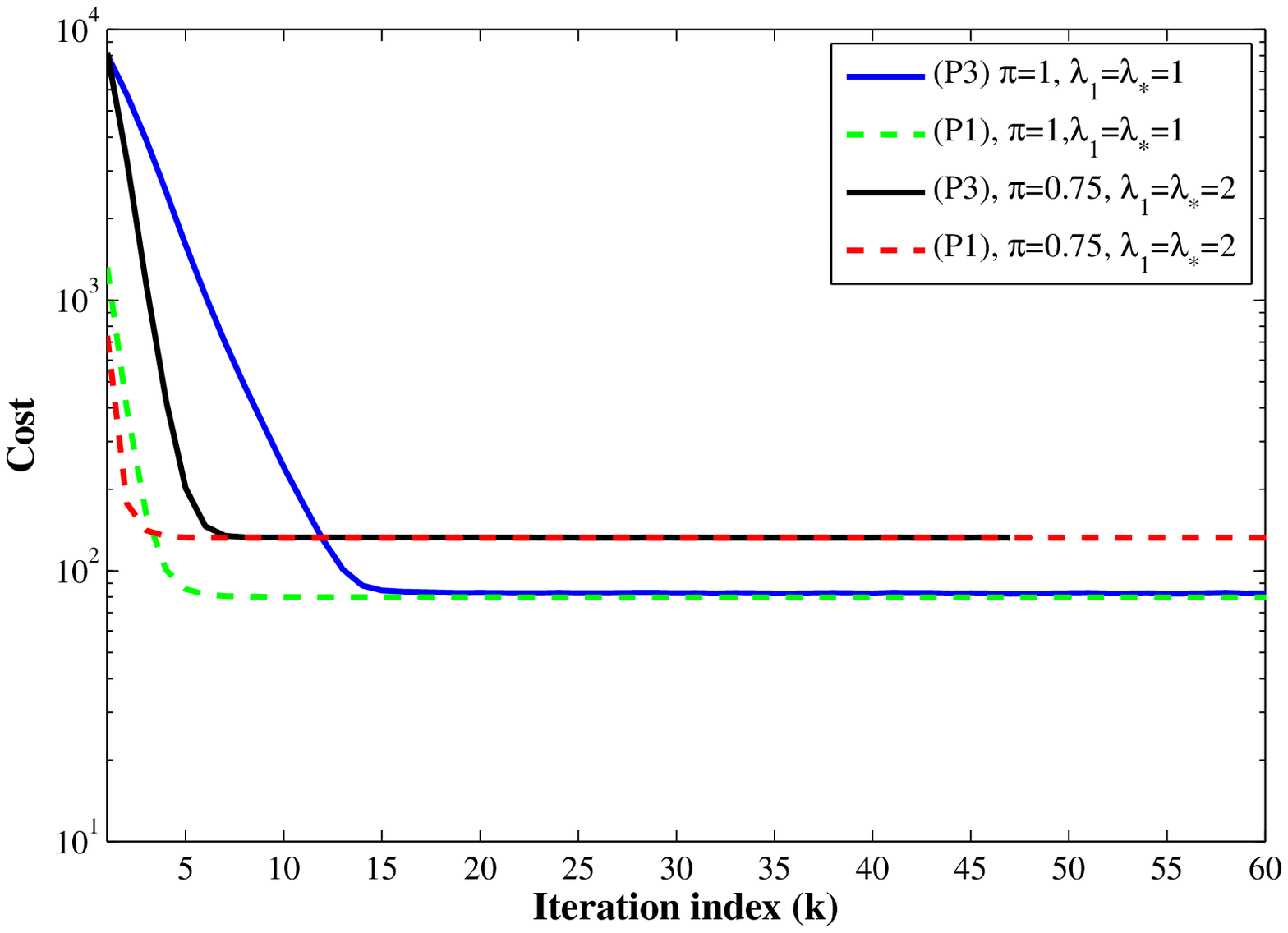,width=0.5
     \linewidth, height=2.3 in } &
     \epsfig{file=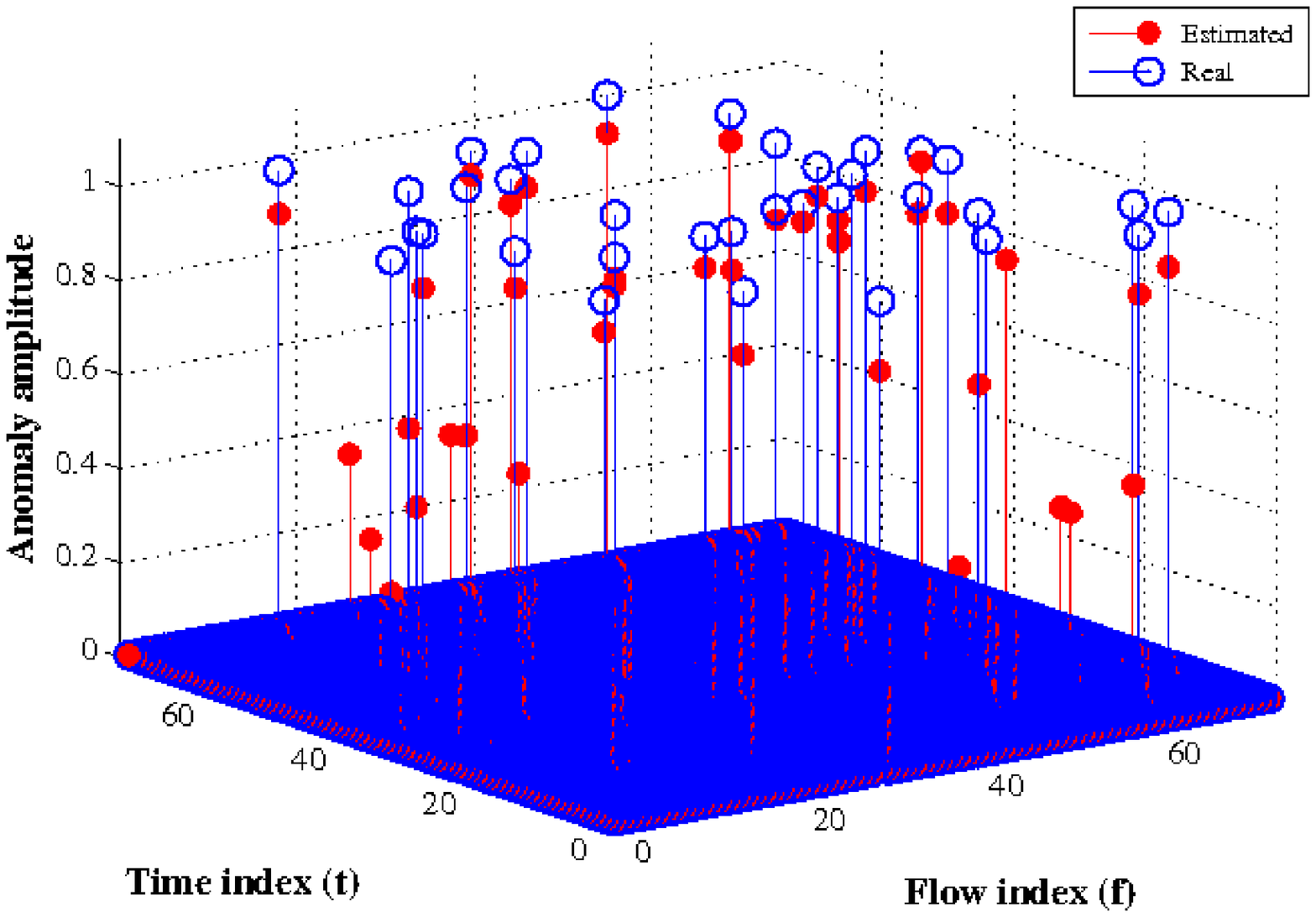,width=0.5
     \linewidth, height=2.3 in } \\
     (a) &
     (b) \\
  \end{tabular}
  \caption{Performance of the batch estimator (P3), for 
  $\sigma=10^{-2}$ and $p=0.005$. (a) Cost of the 
  estimators (P1) and (P3), versus iteration index. (b) Amplitude 
  of the true and estimated anomalies for $\pi=1$ (no missing data), when 
  $P_{\rm FA}=0.0011$ and $P_{\rm D}=0.947$. }
  \label{fig:fig_cnvg_cost_batch}
\end{figure}

\noindent\textbf{Performance of the batch estimator}. To demonstrate the
merits of the batch BCD algorithm for unveiling network anomalies (Algorithm~\ref{tab:table_1}), 
simulated data are generated for a time interval of size $T=100$. For validation purposes, 
the benchmark estimator (P1) is iteratively solved by alternating minimization over  
$\bA$ (which corresponds to Lasso) and $\bX$. The minimizations with respect to $\bX$
can be carried out using the iterative singular-value thresholding (SVT) algorithm~\cite{CCS08}. 
Note that with full data, SVT requires only a single SVD computation. In the presence of 
missing data however, the SVT algorithm may require several SVD computations until
convergence, rendering the said algorithm prohibitively complex for large-scale problems. In contrast, 
Algorithm \ref{tab:table_1} only requires simple $\rho \times \rho$ 
inversions. Fig.~\ref{fig:fig_cnvg_cost_batch} (a) depicts the convergence of
the respective algorithms used to solve (P1) and (P3), for different amounts of missing data 
(controlled by $\pi$). It is apparent that both estimators attain identical
performance after a few tens of iterations, as asserted by Proposition \ref{prop:prop_1}.
To corroborate the effectiveness of Algorithm \ref{tab:table_1} in unveiling
network anomalies across flows and time,
Fig.~\ref{fig:fig_cnvg_cost_batch} (b) maps out the magnitude of the true and
estimated anomalies when $\sigma=10^{-2}$ and $\pi=1$. To discard spurious estimates, consider the
hypothesis test $\hat{a}_{f,t}  \gtreqless_{\mathcal{H}_0}^{\mathcal{H}_1}
0.1$, with anomalous and anomaly-free hypotheses
$\mathcal{H}_1$ and $\mathcal{H}_0$, respectively. The 
false alarm and detection rates achieved are then $P_{\rm FA}=0.0011$ and $P_{\rm D}=0.947$, 
respectively. 

\begin{figure}[t]
\centering
\begin{tabular}{cc}
     \epsfig{file=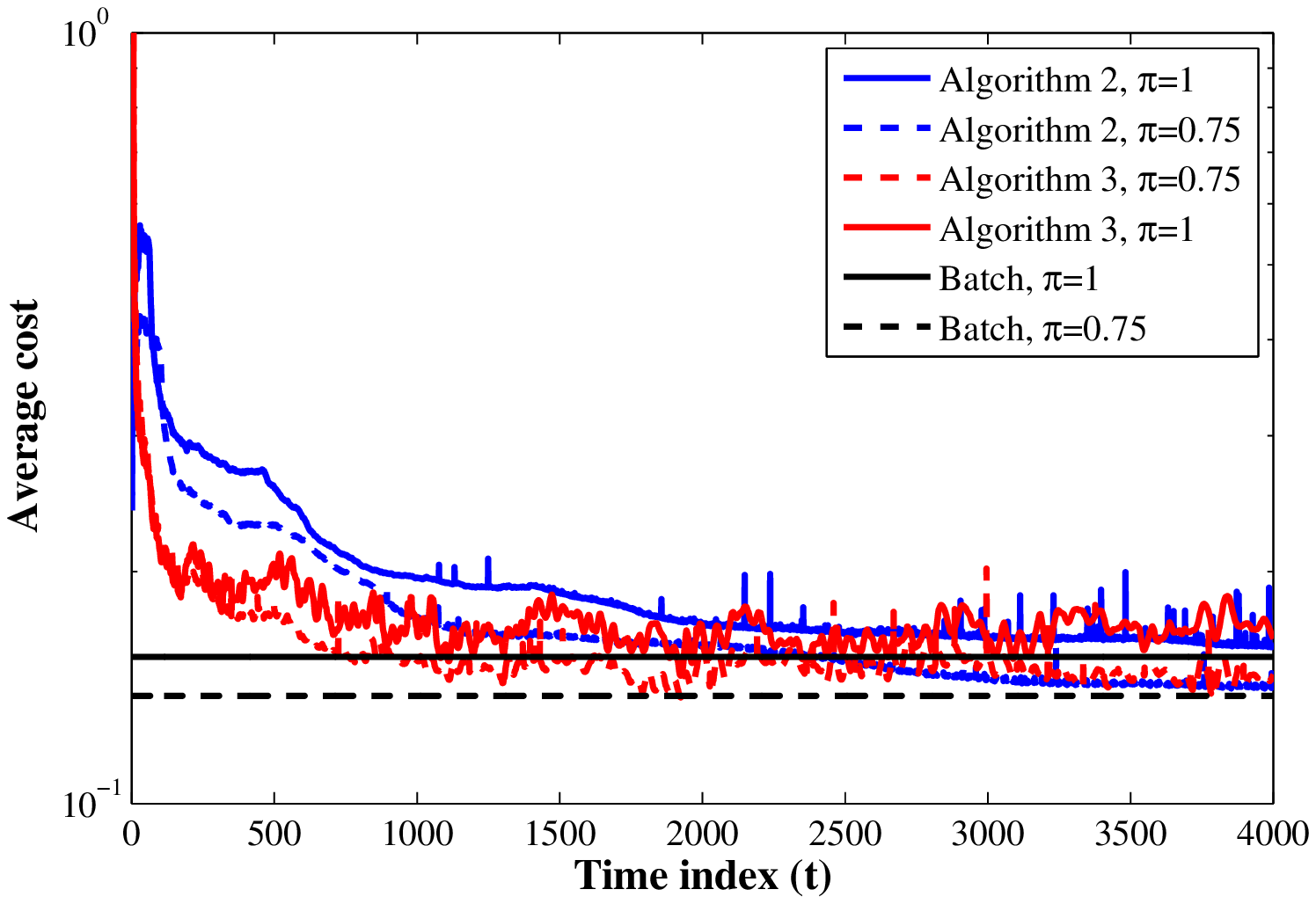,width=0.5\linewidth
     ,height=2.3 in } &
   \epsfig{file=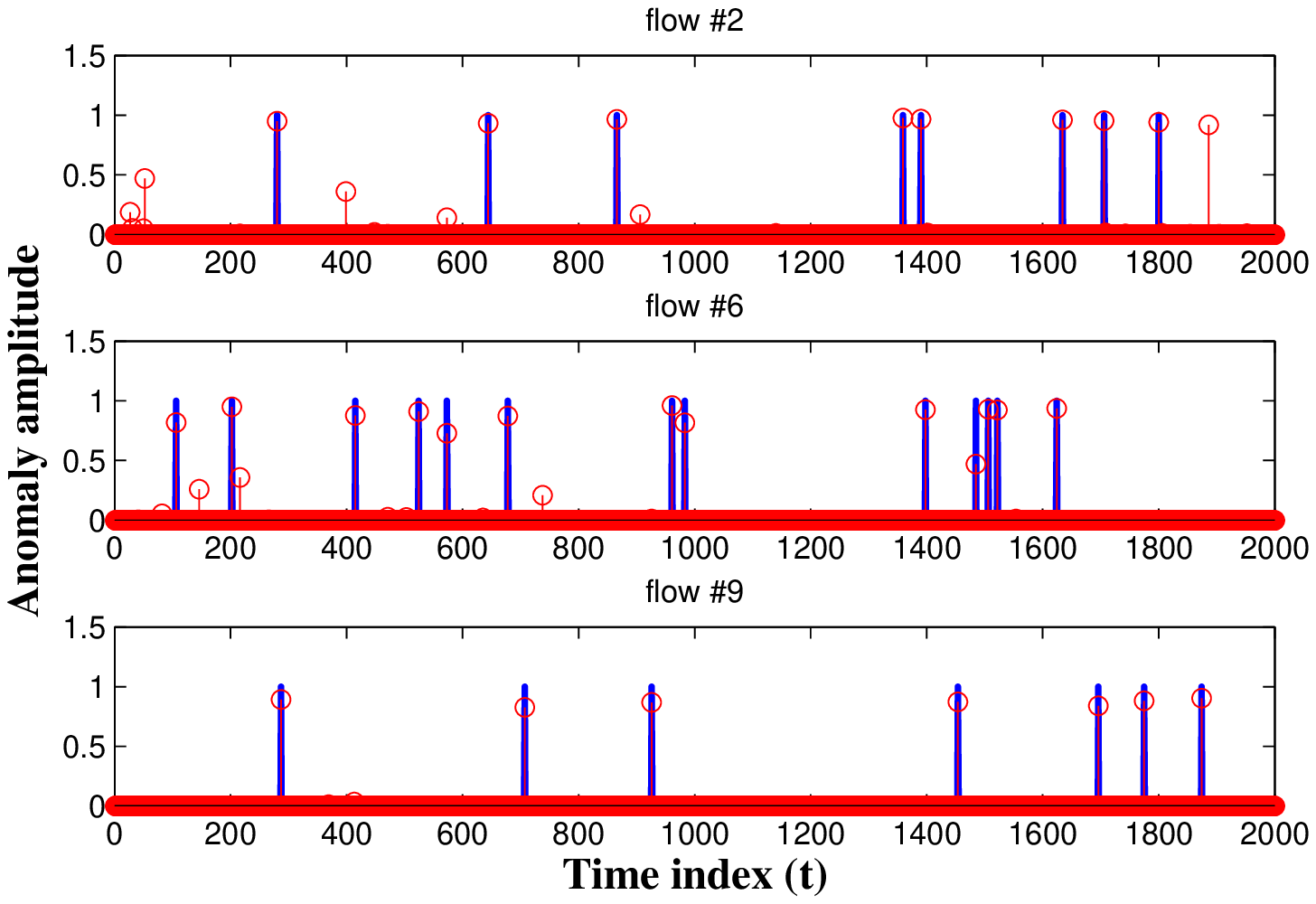,width=0.5
   \linewidth,height=2.3 in } \\
     (a) &
     (b) \\
  \end{tabular}
\caption{Performance of the online estimator for $\sigma=10^{-2}$,
$p=0.005$, $\lambda_1=0.11$, and $\lambda_*=0.36$. (a) Evolution of the average cost $C_t(\bL[t])$ of
the online algorithms versus the batch counterpart (P3). (b) Amplitude of true (solid) 
and estimated (circle markers) anomalies via the
online Algorithm~\ref{tab:table_2}, for three representative flows when $\pi=1$ (no missing data).}

  \label{fig:fig_cnvg_cost_online}
\end{figure}

\noindent\textbf{Performance of the online algorithms}. To confirm the
convergence and effectiveness of the online Algorithms~\ref{tab:table_2} 
and~\ref{tab:table_4}, simulation tests are carried out for 
infinite memory $\beta=1$ and invariant routing matrix $\bR$. 
Figure~\ref{fig:fig_cnvg_cost_online} (a)
depicts the evolutions of the average cost $C_t(\bL_t)$ 
in~\eqref{eq:cost_target} for different
amounts of missing data $\pi=0.75,1$ when the noise level is $\sigma=10^{-2}$. 
It is evident that for both online algorithms the average cost converges (possibly within
a ball) to its batch counterpart in (P3) normalized by the window size $T=t$. 
Impressively, this observation together with the one in 
Fig.~\ref{fig:fig_cnvg_cost_batch} (a) corroborate that 
the online estimators can attain the performance of the benchmark estimator, 
whose stable/exact recovery performance is well documented e.g., 
in~\cite{tit_exactrecovery_2012,zlwcm10,CR08}. It is further observed that the 
more data are missing, the more
time it takes to learn the low-rank nominal traffic subspace, which in turn
slows down convergence.

To examine the tracking capability of the online estimators,~Fig.~\ref{fig:fig_cnvg_cost_online} (b)
depicts the estimated versus
true anomalies over time as Algorithm \ref{tab:table_2} evolves for three
representative flows indicated on Fig.~\ref{fig:fig_nettolpology}, 
namely $f_2,f_6,f_9$ corresponding to the $f=2,6,9$-th rows 
of $\bA_0$. Setting the detection threshold to the value $0.1$ as before, for 
the flows $f_2,f_6,f_9$ Algorithm \ref{tab:table_2} attains detection 
rate $P_{\rm D}=0.83,1,1$ at false alarm rate $P_{\rm 
{FA}}=0.0171,0.0040,0.0081$, respectively. The quantification error 
per flow is also around $P_{\rm Q}=0.7606,0.5863,0.4028$, respectively. 
As expected, more false alarms are declared at early iterations as 
the low-rank subspace has not been learnt accurately. Upon learning the 
subspace performance improves and almost all anomalies are identified. 
Careful inspection of Fig.~\ref{fig:fig_cnvg_cost_online} (b) reveals that
the anomalies for $f_9$ are better identified visually than those for $f_2$. As 
shown in Fig.~\ref{fig:fig_nettolpology}, $f_2$ is carried 
over links $(1,2),(2,4),(4,14),(14,3)$ each one carrying $33,31,35,22$ additional flows, respectively, 
whereas $f_9$ is aggregated over link $(1,3)$ with only $2$ additional flows. Hence,
identifying $f_2$'s anomalies from the highly-superimposed load of links 
$(1,2),(2,4),(4,14),(14,3)$ is a more challenging task relative to link 
$(1,3)$. This simple example manifests the fact that
the detection performance strongly depends on the network topology and the 
routing policy implemented, which determine the routing matrix. In 
accordance with~\cite{tit_exactrecovery_2012}, the coherence of sparse column 
subsets of the routing matrix plays an important role in identifying the 
anomalies. In essence, the more incoherent the column subsets of $\bR$ are, the 
better recovery performance one can attain. An intriguing question left here 
to address in future research pertains to desirable network topologies giving
rise to incoherent routing matrices.

\noindent\textbf{Tracking routing changes}. The measurement model 
in~\eqref{eq:y_t_dynamic} has two time-varying attributes which challenge the identification 
of anomalies. The first one is missing measurement data arising from e.g., packet
losses during the data collection process, and the second one pertains to routing changes due
to e.g., network congestion or link failures. It is thus
important to test whether the proposed online algorithm succeeds in tracking these changes. As
discussed earlier, missing data are sampled 
uniformly at random. To assess the impact of routing
changes on the recovery performance, a simple probabilistic model is
adopted where each time instant a single link fails, or,
returns to the operational state. Let $\bPhi$
denote the adjacency matrix of the network graph $G$, where $[\bPhi]_{i,j} =1$ 
if there exists a physical link joining nodes $i$ and $j$, and zero otherwise. 
Similarly, the active links involved in routing the data at time $t$ 
are represented by the effective adjacency
matrix $\bPhi_t^{\rm {eff}}$. At time instant $t+1$, a biased coin is tossed with 
small success probability $\alpha$, and one of the links, say~$(i,j) \in \bPhi_t^{\rm
{eff}}$, is chosen uniformly at random and removed from $G$ while ensuring that the
network remains connected. Likewise, an edge $(\ell,k)\in\bPhi\backslash
\bPhi_t^{\rm {eff}}$ is added with the same
probability $\alpha$. The resulting adjacency matrix is then $\bPhi_{t+1}^{\rm
{eff}}=\bPhi_t^{\rm  {eff}} + \ind_{\{b_{1,t}\}}\be_{\ell}\be'_k - \ind_{\{b_{1,t}\}}\be_i\be'_j$, 
where the indicator function $\ind_{\{x\in\mathcal{X}\}}$ equals one when
$x\in\mathcal{X}$, and zero otherwise; and $b_{1,t},b_{2,t}\sim\textrm{Ber}(\alpha)$ 
are i.i.d. Bernoulli random variables. The minimum 
hop-count algorithm is then applied to
$\bPhi_{t+1}^{\rm  {eff}}$, to update the routing matrix $\bR_{t+1}$. Note that
$\bR_{t+1}=\bR_{t}$ with probability $(1-\alpha)^2$.

\begin{figure}[t]
\centering
\begin{tabular}{cc}
     \epsfig{file=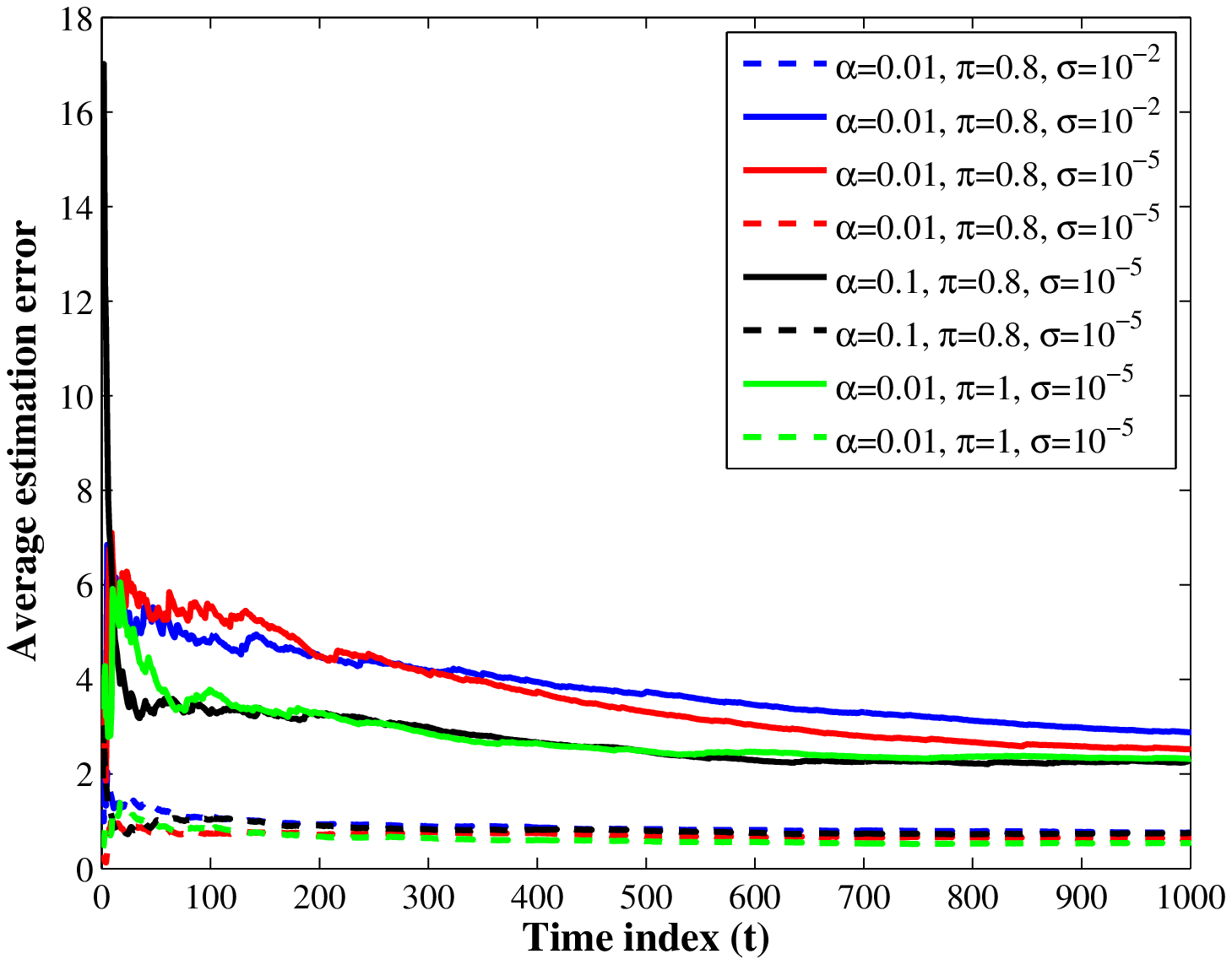,width=0.5
     \linewidth, height=2.3 in } &
     \epsfig{file=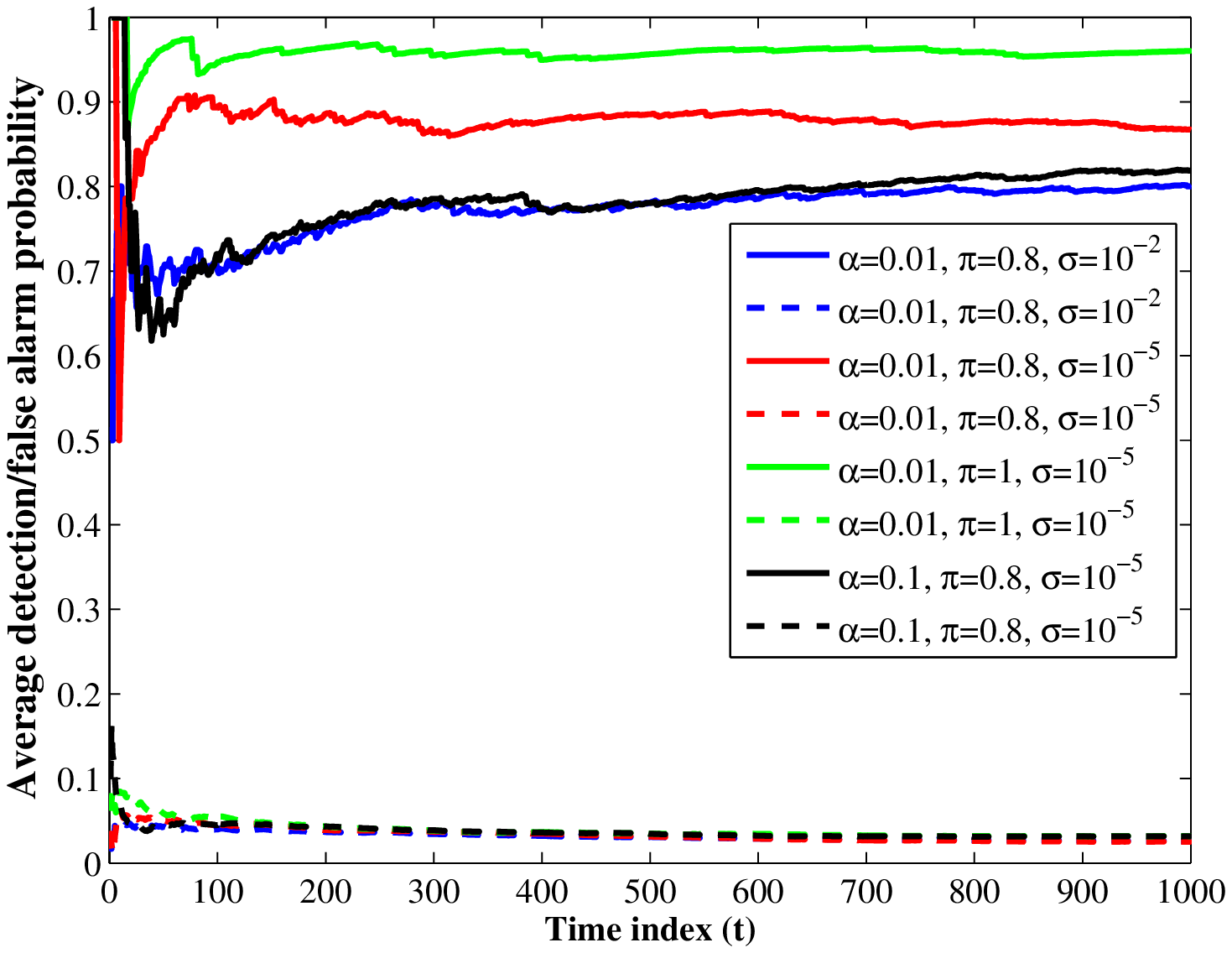,width=0.5
     \linewidth, height=2.3 in } \\
     (a) &
     (b) \\
     \epsfig{file=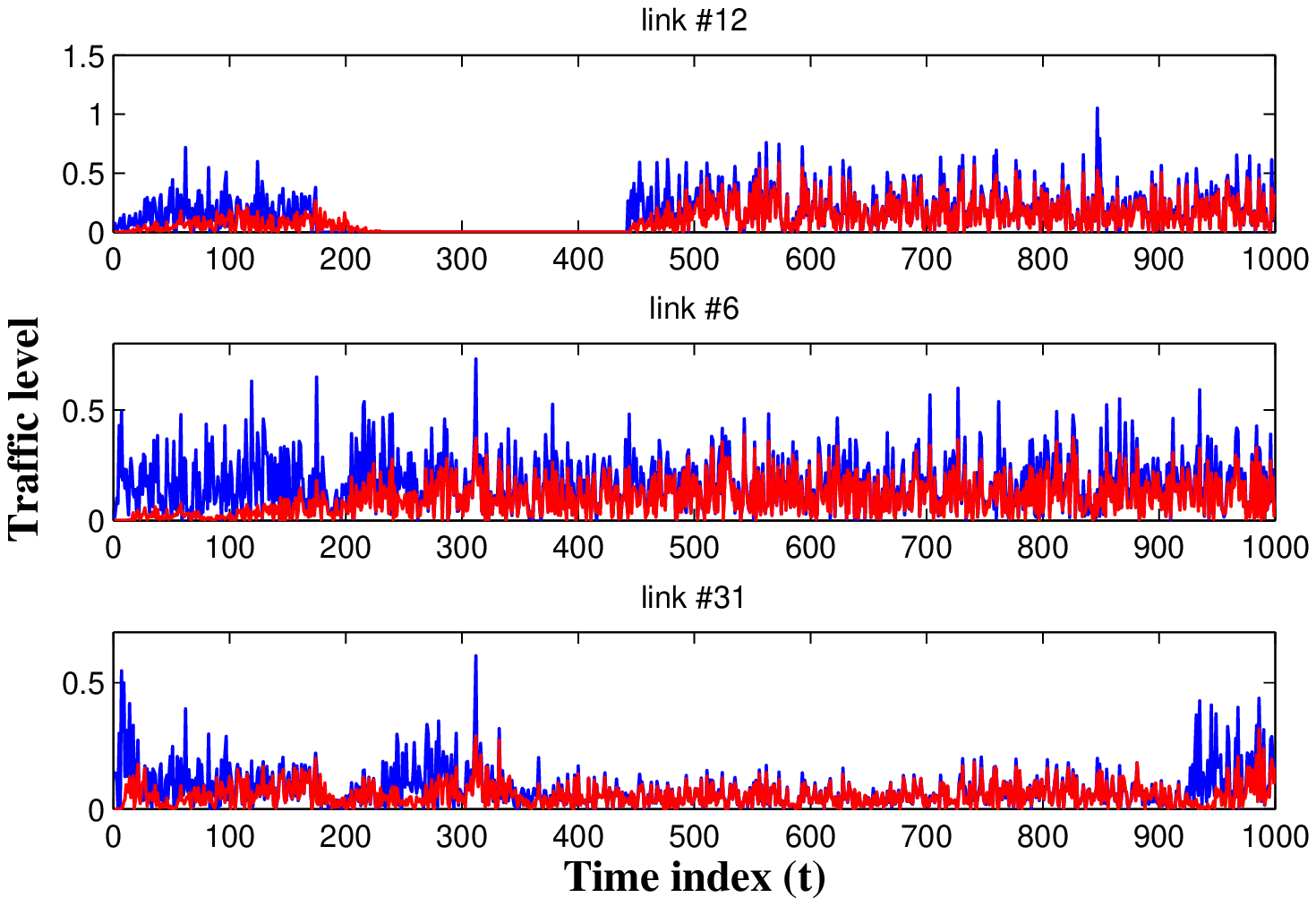,width=0.5
     \linewidth, height=2.3 in } &
     \epsfig{file=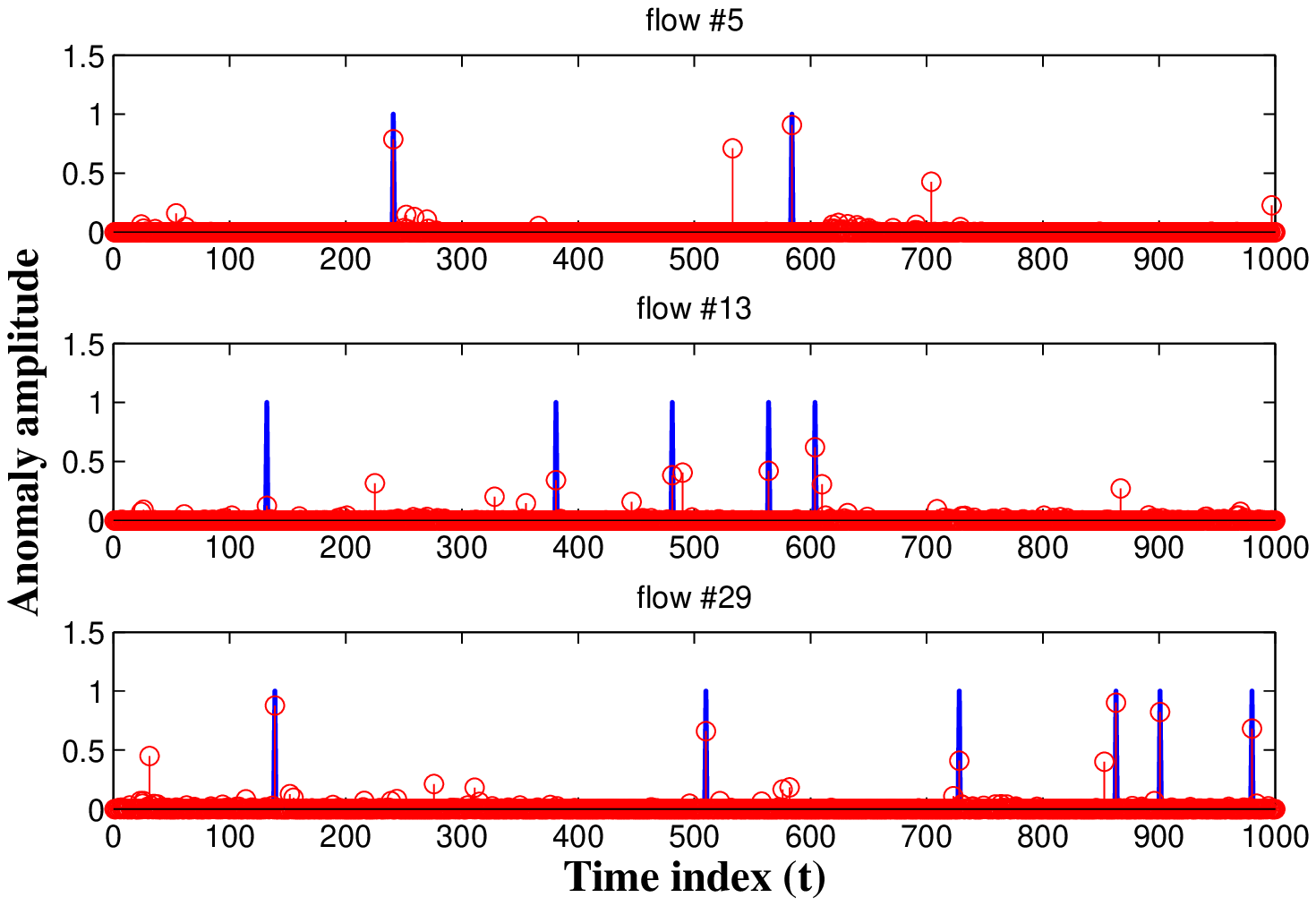,width=0.5
     \linewidth, height=2.3 in } \\
     (c) &
     (d) \\
  \end{tabular}
  \caption{Tracking routing changes for $p=0.005$. (a)
Evolution
of average anomaly (dotted) and traffic (solid) estimation errors. (b)
Evolution of average
detection (solid) and false alarm (dotted) rates. (c) Estimated (red)
versus
true (blue) link
traffic for three representative links. (d) Estimated (circle markers)
versus true (solid) anomalies for three representative flows
when $\pi=0.8$, $\sigma=10^{-5}$, and $\alpha=0.01$. }
  \label{fig:fig_perf_tracking}
\end{figure}

The performance is tested here for fast and slowly varying
routing corresponding to $\alpha=0.1$ and $\alpha=0.01$,
respectively, when $\beta=0.9$. A metric of interest is the average square 
error in estimating the anomalies, namely $e_t^a:=\frac{1}{t}\sum_{i=1}^t\|\hat{\ba}_t -
\ba_t\|_2^2$, and the link traffic, namely $e_t^x:=\frac{1}{t}\sum_{i=1}^t\|\hat{\bx}_t - \bx_t\|_2^2$.
Fig.~\ref{fig:fig_perf_tracking} (a) plots the average estimation error for
various noise variances and amounts of missing data. The estimation error decreases
quickly and after learning the subspace it becomes almost invariant. 
To evaluate 
the support recovery performance of the online estimator, define the 
average detection and false alarm rate
\begin{align}
P_{\rm D}:= \frac{\sum_{\tau=1}^t\sum_{f
=1}^F \ind_{\{\hat{a}_{f,\tau} \geq 0.1, a_{f,\tau} \geq
0.1\}}}{\sum_{\tau=1}^t\sum_{f=1}^F \ind_{\{a_{f,\tau} \geq 0.1\}}}, \hspace{12mm}
P_{\rm FA}:= \frac{\sum_{\tau=1}^t \sum_{f=1}^F \ind_{\{\hat{a}_{f,\tau} \geq 0.1,
a_{f,\tau} \leq0.1\}}}{\sum_{\tau=1}^t \sum_{f=1}^F \ind_{\{a_{f,\tau} \leq 0.1\}}}.
\label{eq:pd_f}
\end{align}
Inspecting~Fig.~\ref{fig:fig_perf_tracking} (b) 
one observes that for $\alpha=0.01$ and $\pi=0.8$, increasing the noise 
variance from $10^{-5}$ to $10^{-2}$ lowers the detection probability by 
$10\%$. Moreover, when $\sigma=10^{-5}$ and $\alpha=0.01$, dropping $20\%$ 
of the observations renders the estimator misdetect $11\%$ more anomalies. The 
routing changes from $\alpha=0.01$ to $\alpha=0.1$ when $\sigma=10^{-5}$ and 
$\pi=0.8$ comes with an adverse effect of about $6\%$ detection-rate decrease. 
For a few representative network links and flows
Fig.~\ref{fig:fig_perf_tracking} (c) and (d) illustrate how Algorithm \ref{tab:table_2} tracks 
the anomalies and link-level traffic. Note that in 
Fig.~\ref{fig:fig_perf_tracking} (c) link $12$ is dropped for the time period 
$t\in[220,420]$, and thus the traffic level becomes zero. The flows being carried 
over link~$31$ are also varying due to routing changes, which occur at
time instants $t={220,940}$ when the traffic is not tracked accurately.


\begin{figure}[t]
\centering
  \centerline{\epsfig{figure=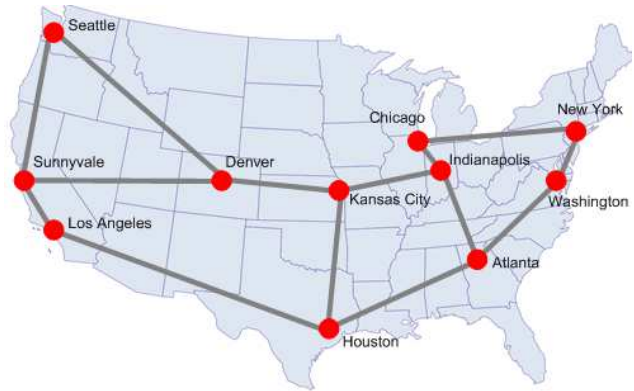,width=0.5\textwidth}}
\vspace{-5mm}\caption{Internet-2 network topology graph.}
  \label{fig:fig_abilene}
\end{figure}

\subsection{Real network data tests}
\label{ssec:perf_real}

\noindent\textbf{Internet-2 network example}. Real data including OD flow
traffic levels are collected from the operation of the Internet-2 network
(Internet backbone network across USA)~\cite{Internet2}, shown in 
Fig.~\ref{fig:fig_abilene}. Flow traffic
levels are recorded every $5$-minute intervals, for a three-week operational
period of Internet-2 during Dec. 8--28,
2008~\cite{Internet2}. Internet-2 comprises $N=11$ nodes,
$L=41$ links, and $F=121$ flows. Given the OD flow traffic measurements, the
link loads in $\bY$ are obtained through multiplication with the Internet-2
routing matrix, which in this case remains invariant during the three weeks of data
acquisition~\cite{Internet2}. Even though $\bY$ is ``constructed'' here from
flow measurements, link loads can be typically acquired from SNMP traces~\cite{MC03}.

The available OD flows are incomplete due to problems in the data collection process. In
addition, flows can be modeled as the superposition of ``clean'' plus anomalous traffic, i.e.,
the sum of some unknown ``ground-truth'' low-rank and sparse matrices
$\cP_{\Omega}(\bX_{0}+\bA_0)$. Therefore, setting $\bR=\bI_F$ in (P1)
one can first run the batch Algorithm~\ref{tab:table_1} 
to estimate the ``ground-truth'' components $\{\bX_0,\bA_0\}$. The estimated
$\bX_0$ exhibits three dominant singular values, confirming the low-rank
property of the nominal traffic matrix. To 
be on the conservative side, only important spikes with magnitude greater
than the threshold level $50\|\bY\|_F/LT$ are retained as benchmark 
anomalies (nonzero entries in $\bA_0$).

\noindent\textbf{Comparison with PCA-based batch 
estimators~\cite{LCD04},~\cite{zggr05}.} To highlight the merits of the batch estimator
(P3), its performance is compared with the spatial 
PCA-based schemes reported in~\cite{LCD04} and~\cite{zggr05}. These methods
capitalize on the fact that the anomaly-free traffic matrix has 
low-rank, while the presence of anomalies considerably increases the rank of 
$\bY$. Both algorithms rely on a two-step estimation procedure: (s1) perform 
PCA on the data $\bY$ to extract the (low-rank) anomaly-free link 
traffic matrix $\tilde{\bX}$; and (s2) declare anomalies based on the 
residual traffic $\tilde{\bY}:=\bY-\tilde{\bX}$. The algorithms in~\cite{zggr05} 
and~\cite{LCD04} differ in the way (s2) is performed. On its operational
phase, the algorithm in~\cite{LCD04} declares the presence of an anomaly at time $t$,
when the projection of $\by_t$ onto the anomalous subspace exceeds a prescribed threshold. 
It is clear that the aforementioned method is unable to identify anomalous flows. 
On the other hand, the network anomography approach of~\cite{zggr05} capitalizes 
on the sparsity of anomalies, and recovers the anomaly matrix by minimizing  $\|\tilde{\bA}\|_1$,
subject to the linear constraints $\tilde{\bY}=\bR\tilde{\bA}$. 

\begin{figure}[t]
\centering
\begin{tabular}{cc}
     \epsfig{file=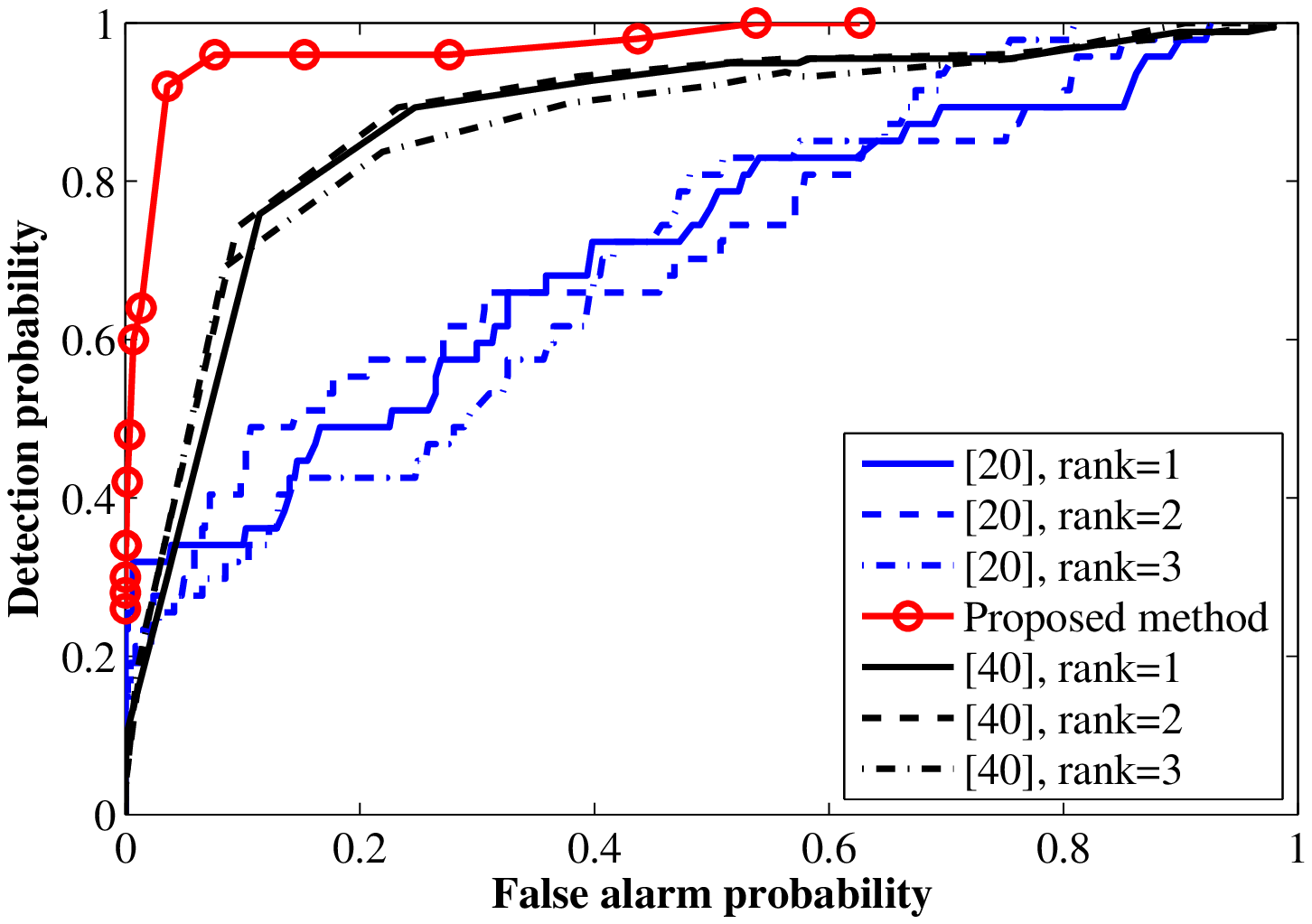,width=0.5
     \linewidth, height=2.3 in } &
     \epsfig{file=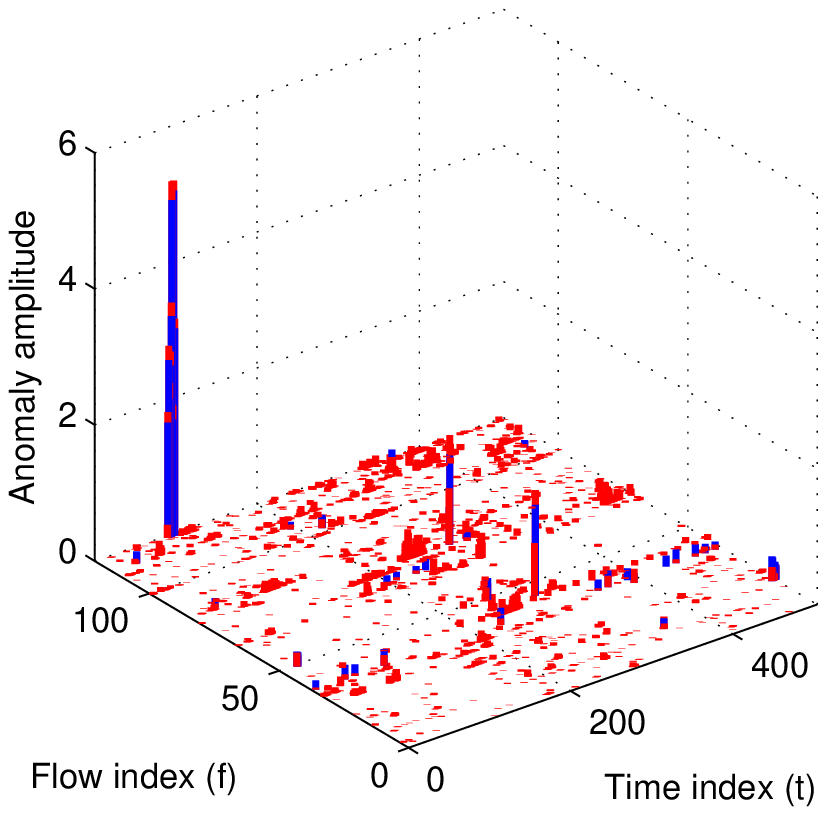,width=0.5
     \linewidth, height=2.3 in } \\
     (a) &
     (b) \\
  \end{tabular}
  \caption{Performance of the batch estimator for Internet-2 network data. (a) 
  ROC curves of the 
  proposed
  versus the PCA-based methods. (b)
  Amplitude of the true (blue) and estimated (red) anomalies for $P_{\rm FA}=0.04$ and 
  $P_{\rm D}=0.93$.}
  \label{fig:fig_batch_pca_real}
\end{figure}

The aforementioned methods require a priori knowledge on the rank of the 
anomaly-free traffic matrix, and assume there is no missing data. 
To carry out performance comparisons, the detection rate will be adopted as figure of
merit, which measures the algorithm's success in identifying anomalies across
both flows and time instants. ROC curves are depicted in Fig. 
\ref{fig:fig_batch_pca_real} (a), for different
values of the rank required to run the PCA-based methods. It is apparent
that the estimator (P3) obtained via Algorithm \ref{tab:table_1} 
markedly outperforms both PCA-based methods in terms of 
detection performance. This is somehow expected, since (P3) advocates
joint estimation of the anomalies and the nominal traffic matrix. For 
an instance of $P_{\rm FA}=0.04$ and $P_{\rm 
D}=0.93$, Fig. \ref{fig:fig_batch_pca_real} (b) illustrates the effectiveness 
of the proposed 
algorithm in terms of unveiling the anomalous flows and time instants.

\begin{figure}[t]
\centering
\begin{tabular}{cc}
\epsfig{file=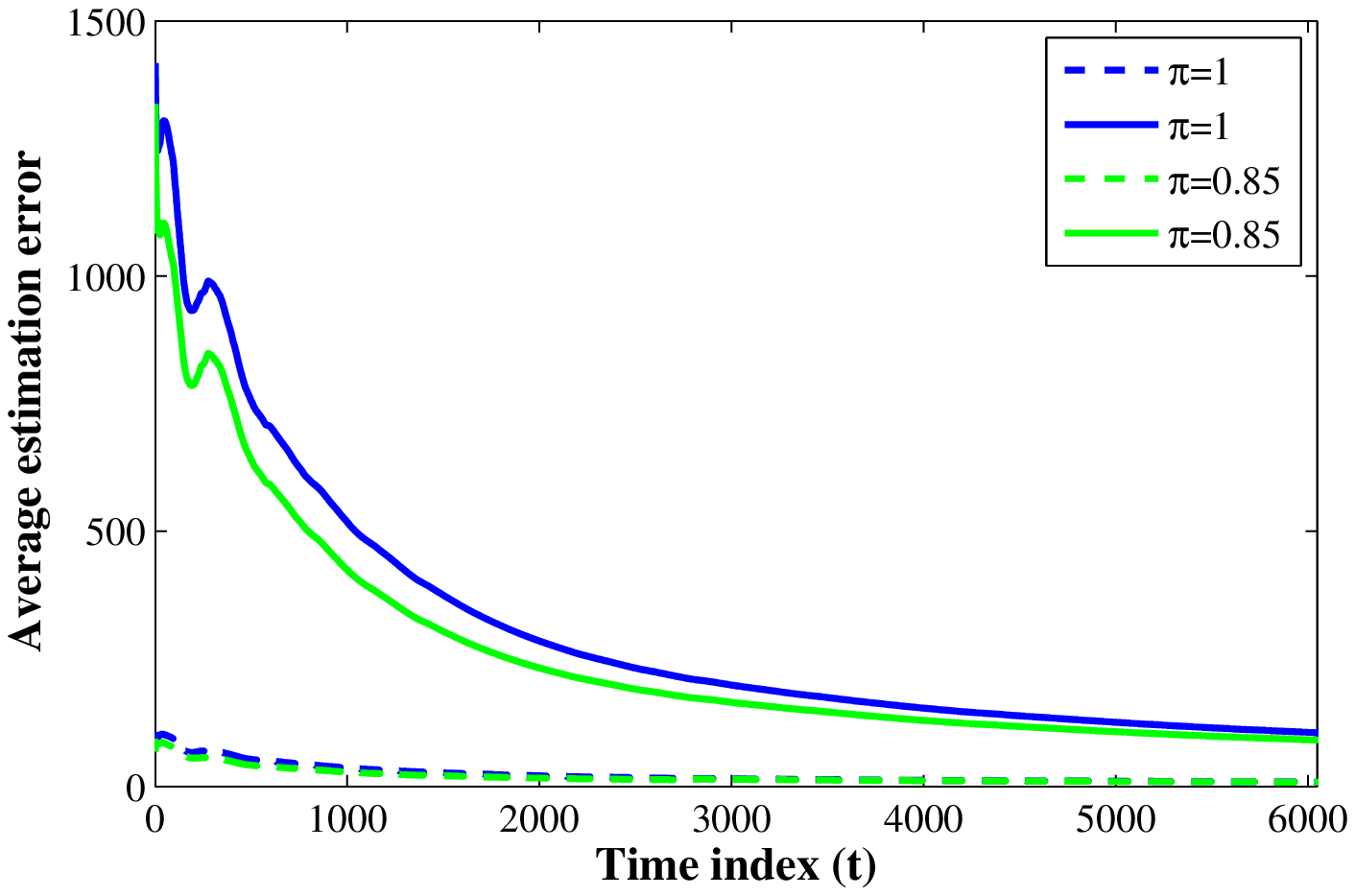,width=0.5
\linewidth, height=2.3 in } &
\epsfig{file=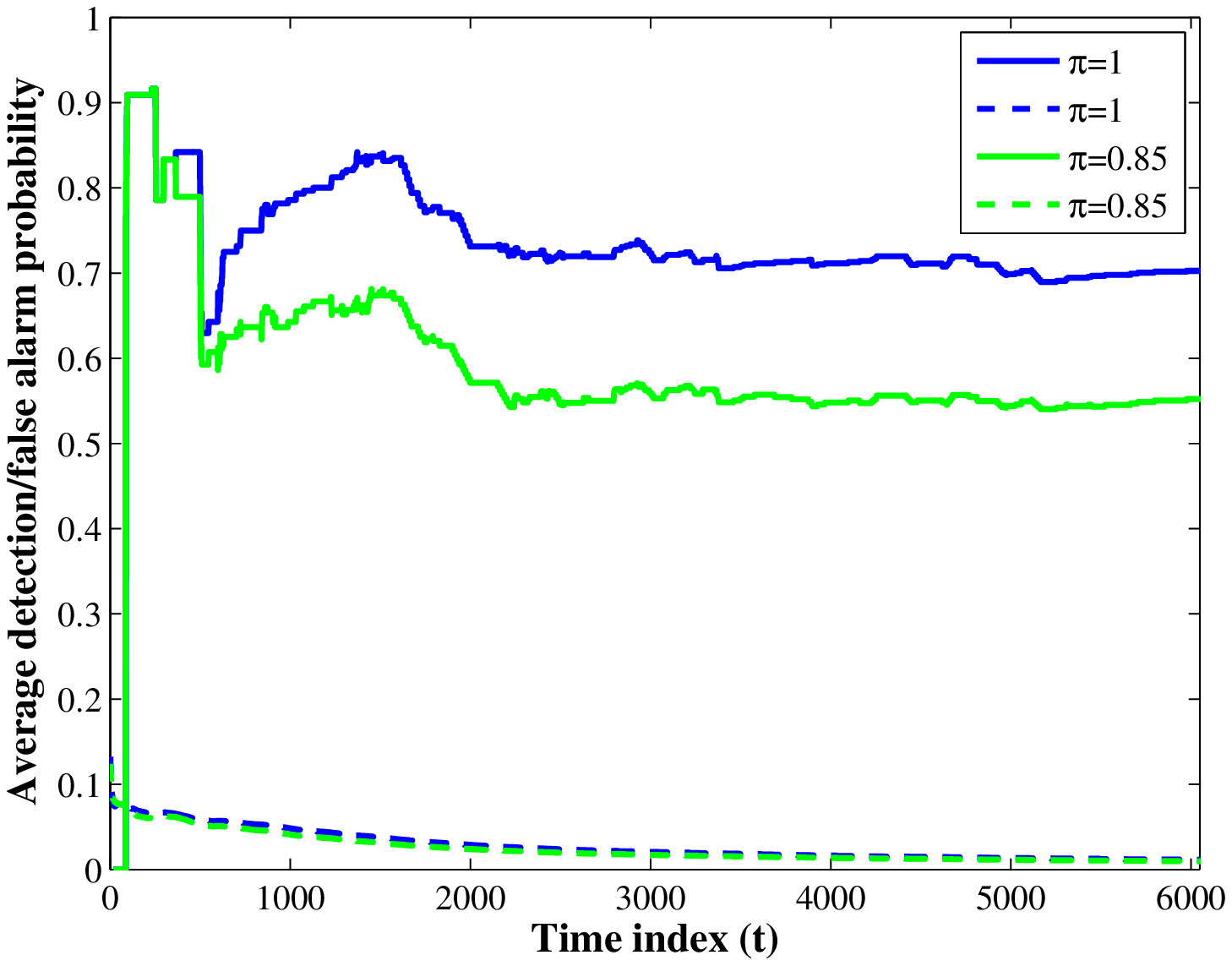,width=0.5
\linewidth, height=2.3 in } \\
(a) &
(b) \\
\epsfig{file=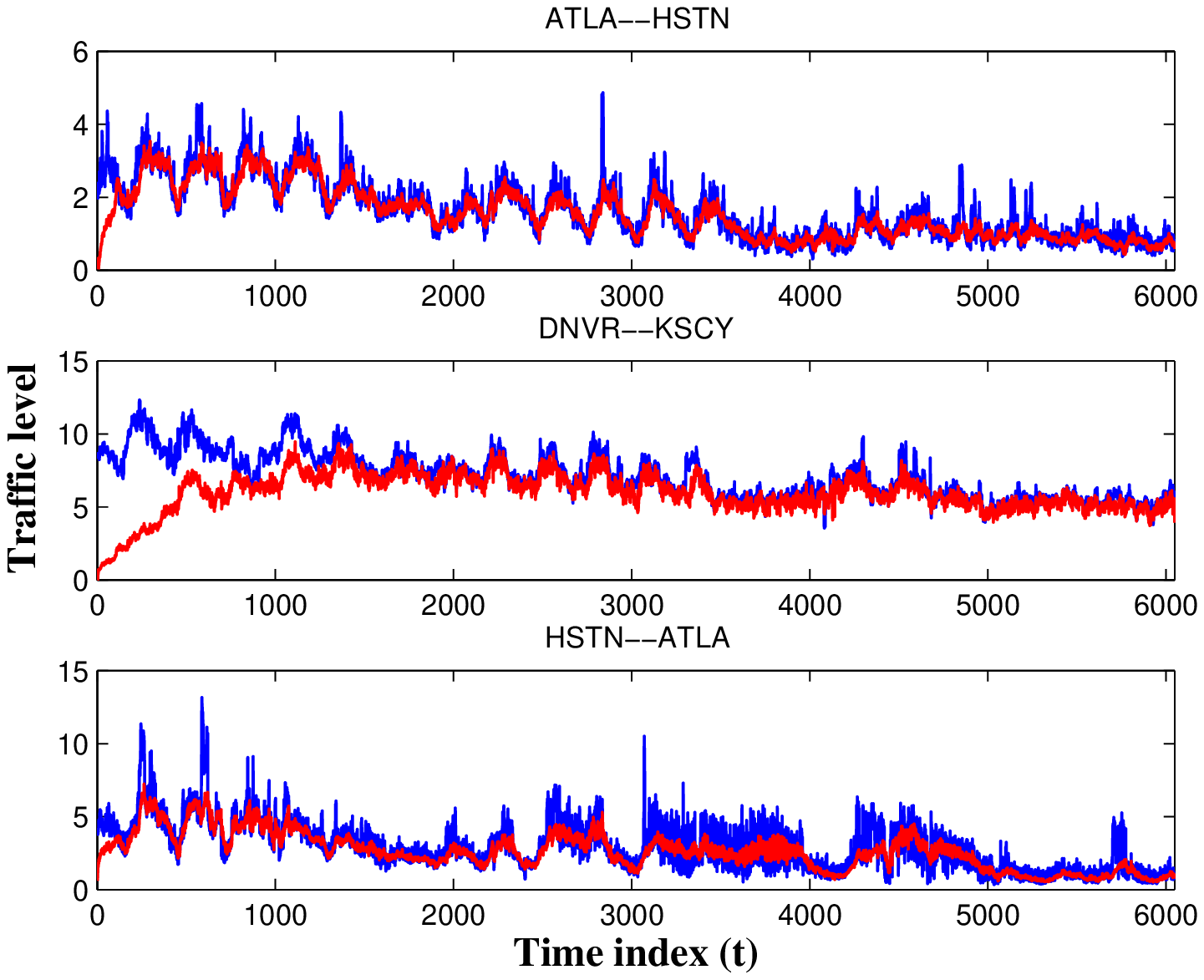,width=0.5
\linewidth, height=2.3 in } &
\epsfig{file=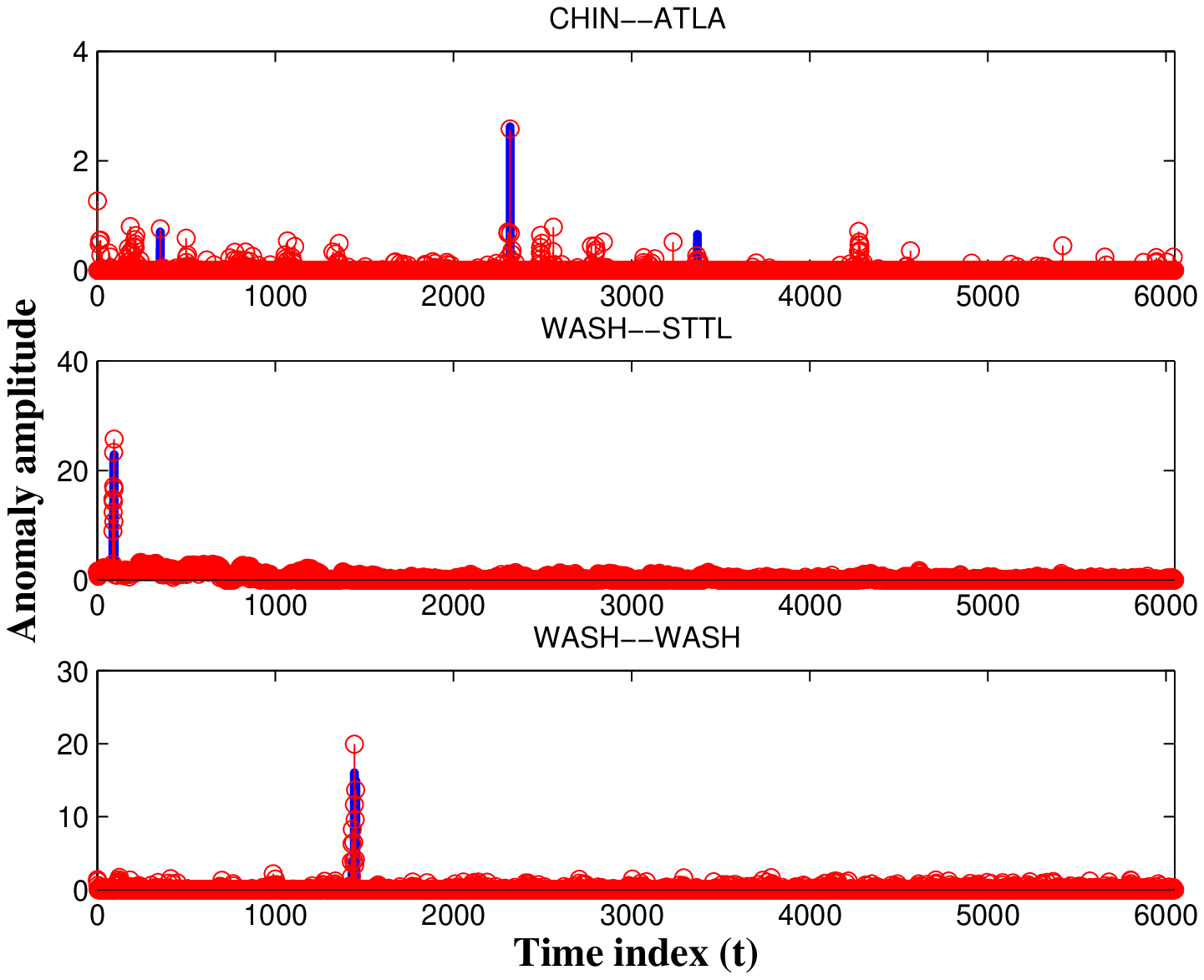,width=0.5
\linewidth, height=2.3 in } \\
(c) &
(d) \\
 \end{tabular}
 \caption{Performance of the online estimator for Internet-2 network data. (a)
 Evolution
 of average anomaly (dotted) and traffic (solid) estimation errors. (b)
 Evolution of average
 detection (solid) and false alarm (dotted) rates. (c) Estimated (red) versus
 true (blue) link
 traffic for thee representative links. (d) Estimated (circle markers)
 versus true (solid) anomalies for three representative flows when $\pi=0.85$.}
 \label{fig:fig_perfonline_realdata}
\end{figure}

\noindent\textbf{Online operation.} Algorithm~\ref{tab:table_2} is
tested here with the Internet-2 network data under two scenarios: with and without missing data. 
For the incomplete data case, a randomly chosen subset of link counts with cardinality $0.15\times LT$ 
is discarded.
The penalty parameters are tuned as $\lambda_1=0.7$ and
$\lambda_{*}=1.4$. The evolution of the average anomaly and traffic estimation
errors, and average detection and false alarm rates are
depicted in Fig.~\ref{fig:fig_perfonline_realdata} (a), (b), respectively. Note
how in the case of full-data, after about a week the 
traffic subspace is accurately learned and the detection (false alarm) 
rates approach the values $0.72$ ($0.011$). It is further observed that even with
$15\%$ missing data, the detection performance degrades gracefully.
Finally, Fig.~\ref{fig:fig_perfonline_realdata}(c)[(d)] depicts how three representative
link traffic levels [OD flow anomalies] are accurately tracked over time.


\section{Concluding remarks}
\label{sec:conc}

An online algorithm is developed in this paper to perform a critical network
monitoring task termed \textit{dynamic anomalography}, meaning
to unveil traffic volume anomalies in backbone networks adaptively. Given link-level
traffic measurements (noisy superpositions of OD flows) acquired sequentially
in time, the goal is to construct a \textit{map} of anomalies in \textit{real time}, that summarizes 
the network `health state' along both the flow and time dimensions. Online algorithms
enable tracking of anomalies in nonstationary environments, typically arising due to to e.g., routing
changes and missing data. The resultant online schemes offer an attractive alternative
to batch algorithms, since they scale gracefully as the number of
flows in the network grows, or, the time window of data acquisition increases.
Comprehensive numerical tests with both synthetic and real network data 
corroborate the effectiveness of the proposed algorithms and their tracking
capabilities, and show that they outperform existing workhorse approaches for
network anomaly detection.


{\Large\appendix}


\noindent\normalsize \emph{\textbf{A. Proof of Lemma~\ref{lem:lemma_2}}}: 
With $\bL_1,\bL_2 \in\mathcal{L}$ consider the function 
\begin{align}
u_t(\ba,\bL_1,\bL_2):=\frac{1}{2} \|\bF_t(\bL_1)(\by_t-\bR_t\ba)\|_2^2-
\frac{1}{2} \|\bF_t(\bL_2)(\by_t-\bR_t\ba)\|_2^2\label{eq:u_t}
\end{align}
where $\bF_t(\bL):=\left[\mathbf{\Omega}_t\left[\bI_L -
\bL\bD_t(\bL)\right]\mathbf{\Omega}_t,~\sqrt{\lambda_{\ast}}\mathbf{\Omega}_t\bD'_t(\bL) \right]'$, and
$\bD_t(\bL):=\left(\lambda_{\ast}\bI_{\rho}+\bL'\mathbf{\Omega}_t\bL
\right)^{-1}\bL'$. It can be readily inferred from a5) that
\begin{align}
u_t(\ba_t(\bL_2),\bL_1,\bL_2) - u_t(\ba_t(\bL_1),\bL_1,\bL_2)
\geq  c_0 \|\ba_t(\bL_2)-\ba_t(\bL_1)\|_2^2 \label{eq:ut_diff_lowbnd}
\end{align}
for some positive constant $c_0$. The rest of the proof deals with Lipschitz
continuity of $u_t(.,\bL_1,\bL_2)$. For $\ba_1$ and $\ba_2$ from a compact set $\mathcal{A}$,
consider
\begin{align}
2|u_t(\ba_1,\bL_1,\bL_2) - 
u_t(\ba_2,\bL_1,\bL_2)|  
 = {}& 2 \langle \bR'_t \left[
 \bF'_t(\bL_2)\bF_t(\bL_2) -
 \bF'_t(\bL_1)
 \bF_t(\bL_1) \right], (\ba_2-\ba_1)\by'_t  \rangle\nonumber\\ 
&  \hspace{-4cm}+\left( \|\bF_t(\bL_1)\bR_t \ba_1\|_2^2 - \|\bF_t(\bL_1)\bR_t \ba_2\|_2^2
\right)- \left( \|\bF_t(\bL_2)\bR_t \ba_1\|_2^2 - \|\bF_t(\bL_2)\bR_t
\ba_2\|_2^2\right).\label{eq:ut_diff_upbnd}
\end{align}
Introducing the auxiliary variable $\bm{\Delta}_a := \ba_2 - \ba_1$, the last
two summands in~\eqref{eq:ut_diff_upbnd} can be bounded as
\begin{align}
 \|\bF_t(\bL_1)\bR_t \ba_1\|_2^2 & - \|\bF_t(\bL_1)\bR_t \ba_2\|_2^2
 -  \|\bF_t(\bL_2)\bR_t \ba_1\|_2^2 + \|\bF_t(\bL_2)\bR_t
\ba_2\|_2^2   \nonumber\\
&\hspace{-1cm} = \left( \|\bF_t(\bL_1)\bR_t \mathbf{\Delta}_a\|_2^2 - \|\bF_t(\bL_2)\bR_t
\mathbf{\Delta}_a\|_2^2 \right) + 2\langle \bR'_t \left[
\bF'_t(\bL_2)\bF_t(\bL_2) - \bF'_t(\bL_1)\bF_t(\bL_1) \right] , \ba_2
\mathbf{\Delta}'_a \ \rangle    \nonumber\\
&\hspace{-1cm} \leq  c_1 \|\bF_t(\bL_2) - \bF_t(\bL_1)\| \|\mathbf{\Delta}_a\|_2^2 + c_2
\|\bF'_t(\bL_2)\bF_t(\bL_2) -
\bF'_t(\bL_1)
\bF_t(\bL_1)\| \|\mathbf{\Delta}_a\|_2 \nonumber\\
&\hspace{-1cm} \leq c_3 \|\bF_t(\bL_2) - \bF_t(\bL_1)\| \|\mathbf{\Delta}_a\|_2
\label{eq:bnd_term1_diff_upbnd}
\end{align}
for some constants $c_1,c_2,c_3 > 0$, since $\|\bF_t(\bL)\|$ for $\bL
\in \mathcal{L}$, $\|\mathbf{\Delta}_a\|_2$, $\|\ba_2\|_2$ for $\ba_1,\ba_2\in
\mathcal{A}$, and $\|\bR_t\|$ are all uniformly bounded. The first summand
on the right-hand side of~\eqref{eq:ut_diff_upbnd} is similarly bounded (details
omitted here). Next, to establish that $\bF_t(\bL)$ is Lipschitz one can derive
the following bound ($\mathbf{\Delta}_L:=\bL_2-\bL_1$)
\begin{align}
\|\bF_t(\bL_2) - \bF_t(\bL_1)\| &\leq  \|\mathbf{\Omega}_t\left[
\bL_2\bD_t(\bL_2) - \bL_1\bD_t(\bL_1)\right]\mathbf{\Omega}_t\| +
\sqrt{\lambda_{\ast}} \|\mathbf{\Omega}_t (\bD'_t(\bL_2) - \bD'_t(\bL_1))\|
\nonumber\\
& \leq \| \bL_1\| (\| \bL_1\| + \sqrt{\lambda_{\ast}} ) \| (\lambda_{\ast}
\bI_{\rho} + \bL'_2
\mathbf{\Omega}_t\bL_2)^{-1} - (\lambda_{\ast} \bI_{\rho} + \bL'_1
\mathbf{\Omega}_t\bL_1)^{-1} \| \nonumber\\
&\hspace{15mm}+  \| \mathbf{\Delta}_L\|
(\|\bL_1\| + \|\bL_2\| + \sqrt{\lambda_{\ast}}) \|(\lambda_{\ast} \bI_{\rho} +
\bL'_2
\mathbf{\Omega}_t\bL_2)^{-1}\|. \label{eq:bnd_diff_Ft}
\end{align}
Upon introducing $\bG'_t\bG_t:=\mathbf{\Delta}'_L  \mathbf{\Omega}_t \bL_1 +
\mathbf{\Delta}'_L  \mathbf{\Omega}_t \mathbf{\Delta}_L + \bL'_1
\mathbf{\Omega}_t \mathbf{\Delta}_L$ and $\bH_t:=\lambda_{\ast}\bI_{\rho}+\bL
\mathbf{\Omega}_t\bL'$ by utilizing the matrix inversion lemma, the first term
is bounded as follows
\begin{align}
 \| (\lambda_{\ast} \bI_{\rho} + \bL'_2
\mathbf{\Omega}_t\bL_2)^{-1} - (\lambda_{\ast} \bI_{\rho} + \bL'_1
\mathbf{\Omega}_t\bL_1)^{-1} \| ={} & \| \bH_t^{-1}(\bL_1) \bG_t (\bI +
\bG'_t\bH_t^{-1}(\bL_1)\bG_t )^{-1} \bG'_t \bH_t^{-1}(\bL_1) \| \nonumber\\
\leq{} & \| \bH_t^{-1}(\bL_1)\|^2 \|\bG_t\|^2 \|(\bI +
\bG'_t\bH_t^{-1}(\bL_1)\bG_t
)^{-1}\|  \nonumber\\
\leq {}& \left(\frac{1}{\lambda_{\ast}}\right)^2 \|\bG_t\|^2  \leq c_4 \|\mathbf{\Delta}_L
\|.
\label{eq:bnd_term1_bndup_diff_Ft}
\end{align}
Putting the pieces together $\bF_t(.)$ is found to be Lipschitz and
subsequently~\eqref{eq:ut_diff_upbnd} is bounded by a constant factor of
$\|\mathbf{\Delta}_L \| \|\mathbf{\Delta}_a \|_2$. Substituting
$\ba_1=\ba_t(\bL_1)$ and $\ba_2=\ba_t(\bL_2)$ along
with the bound in~\eqref{eq:ut_diff_lowbnd} yields the desired result
$\|\ba_t(\bL_2)-\ba_t(\bL_1)\|_2 \leq c_5 \|\bL_2-\bL_1\|$.
Furthermore, from the relationship $\bq_t = \bD_t(\bL)
\mathbf{\Omega}_t(\by_t-\bR_t \ba_t)$, Lipschitz continuity of $\bq_t(\bL)$
readily follows.

Moreover, $g_t(\bL,\bq[t],\ba[t])$ is a quadratic function on a
compact set, and thus clearly Lipschitz continuous. To prove Lipschitz
continuity of $\ell_t(\bL)$, recall the definition
$\{\bq_t(\bL),\ba_t(\bL)\}=\arg\min_{\{\bq,\ba\}} g_t(\bL,\bq,\ba)$ to
obtain after some algebra
\begin{align}
\ell_t(\bL_2) - \ell_t(\bL_1) ={}& \frac{1}{2}
\|\cP_{\Omega_t}(\bL_2\bq_t(\bL_2)
+ \bR_t\ba_t(\bL_2))\|_2^2 - \|\cP_{\Omega_t}(\bL_1\bq_t(\bL_1) +
\bR_t\ba_t(\bL_1))\|_2^2 \nonumber\\& - \langle \cP_{\Omega_t}(\by_t),
\bL_2\bq_t(\bL_2) + \bR_t\ba_t(\bL_2) - \bL_1\bq_t(\bL_1) -
\bR_t\ba_t(\bL_1)    \rangle \nonumber\\
&+\frac{\lambda_{\ast}}{2} \big(\|\bq_t(\bL_2)\|_2^2 -
\|\bq_t(\bL_1)\|_2^2\big) +
\lambda_1
\big(\|\ba_t(\bL_2)\|_1 - \|\ba_t(\bL_1)\|_1\big). \label{eq:l2_l1}
\end{align}
The first term in the right-hand side of \eqref{eq:l2_l1} is bounded as
\begin{align}
&\|\cP_{\Omega_t}(\bL_2\bq_t(\bL_2)
+ \bR_t\ba_t(\bL_2))\|_2^2 - \|\cP_{\Omega_t}(\bL_1\bq_t(\bL_1) +
\bR_t\ba_t(\bL_1))\|_2^2 \leq \nonumber\\
&\hspace{28mm} \big(\|\cP_{\Omega_t}(\bL_2\bq_t(\bL_2) - \bL_1\bq_t(\bL_1))\|_2
+\|\cP_{\Omega_t}(\bR_t\ba_t(\bL_2) - \bR_t\ba_t(\bL_1)) \|_2\big) \nonumber\\
&\hspace{28mm} \times \big(\|\cP_{\Omega_t}(\bL_2\bq_t(\bL_2) +
\bR_t\ba_t(\bL_2))\|_2 +
\|\cP_{\Omega_t}(\bL_1\bq_t(\bL_1) + \bR_t\ba_t(\bL_1)) \|_2\big) \nonumber\\
&\hspace{28mm}\leq c_6 \big( \|\bL_2-\bL_1\| \|\bq_t(\bL_2)\|_2 + \|\bL_1\|
\|\bq_t(\bL_2) -
\bq_t(\bL_1)\|_2 +
\|\bR_t\| \|\ba_t(\bL_2) - \ba_t(\bL_1) \|_2\big) \label{eq:bnd_term1}
\end{align}
for some constant $c_6>0$. The second one is bounded as
\begin{align}
&\langle \cP_{\Omega_t}(\by_t),
\bL_2\bq_t(\bL_2) + \bR_t\ba_t(\bL_2) - \bL_1\bq_t(\bL_1) -
\bR_t\ba_t(\bL_1)    \rangle \nonumber\\
&\hspace{20mm} \leq \|\cP_{\Omega_t}(\by_t)\|_2
\big(\|\cP_{\Omega_t}(\bL_2\bq_t(\bL_2) - \bL_1\bq_t(\bL_1))\|_2 +
\|\cP_{\Omega_t}(\bR_t\ba_t(\bL_2) - \bR_t\ba_t(\bL_1)) \|_2\big) \nonumber\\
&\hspace{20mm} \leq \|\cP_{\Omega_t}(\by_t)\|_2
\big( \|\bL_2-\bL_1\| \|\bq_t(\bL_2)\|_2 + \|\bL_1\| \|\bq_t(\bL_2) -
\bq_t(\bL_1)\|_2 +
\|\bR_t\| \|\ba_t(\bL_2) - \ba_t(\bL_1) \|_2\big). 
\label{eq:bnd_term2}
\end{align}
Finally, one can bound the third term in \eqref{eq:l2_l1} as
\begin{align}
&\frac{\lambda_{\ast}}{2} \big(\|\bq_t(\bL_2)\|_2^2 - \|\bq_t(\bL_1)\|_2^2 \big)+\lambda_1
\big(\|\ba_t(\bL_2)\|_1 - \|\ba_t(\bL_1)\|_1\big) \leq \nonumber \\
&\hspace{30mm}\frac{\lambda_{\ast}}{2} \|\bq_t(\bL_2)-\bq_t(\bL_1)\|_2
\big(\|\bq_t(\bL_2)\|_2+\|\bq_t(\bL_1)\|_2\big) + \lambda_1 \sqrt{F}
\|\ba_t(\bL_2) - \ba_t(\bL_1)\|_2. \label{eq:bnd_term3}
\end{align}
Since $\bq_t(\bL)$ and $\ba_t(\bL)$ are Lipschitz as proved earlier, and
$\bL_1,\bL_2 \in \mathcal{L}$ are uniformly bounded, the expressions in the
right-hand side of \eqref{eq:bnd_term1}-\eqref{eq:bnd_term3} are upper bounded
by a constant factor of $\|\bL_2-\bL_1\|$, and so is $|\ell_t(\bL_2) - \ell_t(\bL_1)|$
after applying the triangle inequality to \eqref{eq:l2_l1}.

Regarding $\nabla \ell_t(\bL)$, notice first that since
$\{\bq_t(\bL),\ba_t(\bL)\}$ is the unique minimizer of $g_t(\bL,\bq,\ba)$ [cf. a5)],
Danskin's theorem~\cite[Prop. B.25(a)]{Bers} implies that
$\nabla \ell_t(\bL) = \cP_{{\Omega}_t}(\by_t - \bL\bq_t(\bL) - \bR_t\ba_t(\bL))
\bq'_t(\bL)$. 
In the sequel, the triangle inequality will be used to split the norm in the right-hand side of
\begin{align}
\big\|\nabla \ell_t(\bL_2)  - \nabla \ell_t(\bL_1)\big\|_F =&
\big\|\cP_{\Omega_t}(\by_t)
\left[\bq_t(\bL_2) - \bq_t(\bL_1)\right]' - \left[
\cP_{\Omega_t}(\bL_2\bq_t(\bL_2))\bq'_t(\bL_2) - \cP_{\Omega_t}(
\bL_1\bq_t(\bL_1))\bq'_t(\bL_1) \right] \nonumber\\
& - \left[
\cP_{\Omega_t}(\bR_t\ba_t(\bL_2))\bq'_t(\bL_2) -
\cP_{\Omega_t}(\bR_t\ba_t(\bL_1))\bq'_t(\bL_1)
\right]\big\|_F.\label{eq:bnd_diff_gradell}
\end{align}
The first term inside the norm is bounded as
\begin{align}
\|\cP_{\Omega_t}(\by_t) \left[\bq_t(\bL_2) - \bq_t(\bL_1)\right]'\|_F \leq
\|\cP_{\Omega_t}(\by_t)\|_2 \|\bq_t(\bL_2) - \bq_t(\bL_1)\|_2.
\label{eq:bnd_term1_grad}
\end{align}
After some algebraic manipulations, the second term is also bounded as
\begin{align}
\|\cP_{\Omega_t}(\bL_2\bq_t(\bL_2))\bq'_t(\bL_2) - \cP_{\Omega_t}(
\bL_1\bq_t(\bL_1))\bq'_t(\bL_1)\|_F \leq{}& \|\bL_2 -
\bL_1\|_F \|\bq_t(\bL_2)\|_2^2\nonumber\\ &\hspace{-2cm}
+ \|\bq_t(\bL_2) - \bq_t(\bL_1)\|_2 \big( \|\bq_t(\bL_2)\|_2 + \| \bq_t(\bL_1)\|_2
\big)  \label{eq:bnd_term2_grad}
\end{align}
and finally one can simply bound the third term as
\begin{align}
\| \cP_{\Omega_t}(\bR_t\ba_t(\bL_2))\bq'_t(\bL_2) -
\cP_{\Omega_t}(\bR_t\ba_t(\bL_1))\bq'_t(\bL_1) \|_F \leq
{}&\|\bR_t\| \big( \|\ba_t(\bL_2)-\ba_t(\bL_1)\|_2 \|\bq_t(\bL_1)\|_2\nonumber\\
&
+\|\bq_t(\bL_2)-\bq_t(\bL_1)\|_2 \|\ba_t(\bL_1)\|_2 \big). \label{eq:bnd_term3_grad}
\end{align}
Since $\ba_t(\bL)$ and $\bq_t(\bL)$ are Lipschitz and uniformly
bounded, from~\eqref{eq:bnd_term1_grad}-\eqref{eq:bnd_term3_grad} one can
easily deduce that $\nabla \ell_t(.)$ is indeed Lipschitz continuous.\hfill$\blacksquare$


\noindent\normalsize \emph{\textbf{B. Proof of Lemma~\ref{lem:lemma_3}}}:
Exploiting that $\nabla \hat{C}_t(\bL[t])=\nabla
\hat{C}_{t+1}(\bL[t+1])=\mathbf{0}_{L\times \rho}$ by algorithmic construction 
and the strong convexity assumption on $\hat{C}_t$ [cf. a4)], 
application of the mean-value theorem readily yields
\begin{align}
&\hat{C}_t(\bL[t+1]) \geq \hat{C}_t(\bL[t]) + \frac{c}{2}
\|\bL[t+1]-\bL[t]\|_F^2 \nonumber\\
&\hat{C}_{t+1}(\bL[t]) \geq \hat{C}_{t+1}(\bL[t+1]) +
\frac{c}{2} \|\bL[t+1]-\bL[t]\|_F^2. \nonumber
\end{align}
Upon defining the function $h_t(\bL):=\hat{C}_{t}(\bL)-\hat{C}_{t+1}(\bL)$
one arrives at
\begin{align}
c\|\bL[t+1]-\bL[t]\|_F^2 \leq h_{t}(\bL[t+1]) - h_t(\bL[t]).  \label{eq:low_bnd_h_diff}
\end{align}
To complete the proof, it suffices to show that
$h_t$ is Lipschitz with constant $\mathcal{O}(1/t)$, and upper bound the right-hand side of 
\eqref{eq:low_bnd_h_diff} accordingly.
Since [cf. \eqref{eq:cost_apx}]
\begin{equation}\label{eq:def_h}
h_t(\bL)=\frac{1}{t(t+1)} \sum_{\tau=1}^t g_\tau(\bL,\bq[\tau],\ba[\tau]) -
\frac{1}{t+1} g_{t+1}(\bL,\bq[t+1],\ba[t+1]) +
\frac{\lambda_{\ast}}{2t(t+1)} \|\bL\|_F^2
\end{equation}
and $g_i(\bL)$ is Lipschitz according to
Lemma~\ref{lem:lemma_2}, it follows that $h_t$ is Lipschitz with constant $\mathcal{O}(1/t)$. 
\hfill$\blacksquare$


\noindent\normalsize \emph{\textbf{C. Proof of Lemma~\ref{lem:lemma_4}}}: 
The first step of the proof is to show that $\{\hat{C}_t(\bL[t])\}_{t=1}^\infty$ 
is a quasi-matringale sequence, and hence convergent a.s.~\cite{Ljung}. Building on the variations of
$\hat{C}_t(\bL[t])$, one can write
\begin{align}
\hat{C}_{t+1}(\bL[t+1]) - \hat{C}_{t}(\bL[t]) {}={}& \hat{C}_{t+1}(\bL[t+1]) -
\hat{C}_{t+1}(\bL[t]) + \hat{C}_{t+1}(\bL[t]) - \hat{C}_{t}(\bL[t])
\nonumber\\
\stackrel{(a)}{\leq} {}& \hat{C}_{t+1}(\bL[t]) - \hat{C}_t(\bL[t])
\nonumber\\
 ={} & \frac{1}{t+1} \left[ g_{t+1}(\bL[t],\bq[t+1],\ba[t+1]) -
 \frac{1}{t} \sum_{\tau=1}^t g_{\tau}(\bL[\tau],\bq[\tau],\ba[\tau])
 \right]\nonumber\\
\stackrel{(b)}{\leq} {}& \frac{1}{t+1} \left[
 g_{t+1}(\bL[t],\bq[t+1],\ba[t+1]) -
 \frac{1}{t} \sum_{\tau=1}^t \ell_{\tau}(\bL[t]) \right]  \label{eq:bnd_variations}
\end{align}
where (a) uses that $ \hat{C}_{t+1}(\bL[t+1]) \leq
\hat{C}_{t+1}(\bL[t])$, and (b) follows from $C_t(\bL[t]) \leq \hat{C}_t(\bL[t])$.

Collect all past data in $\mathcal{F}_t=\{(\Omega_\tau,\by_\tau):\tau \leq
t\}$, and recall that under a1) the random processes $\{\Omega_t,\by_t\}$ are i.i.d. over
time. Then, the expected variations of the approximate cost function are bounded as
\begin{align}
\mathbb{E} \left[ \hat{C}_{t+1}(\bL[t+1]) - \hat{C}_{t}(\bL[t]) |
\mathcal{F}_t \right]  \leq {}& \frac{1}{t+1} \left( \mathbb{E}
[g_{t+1}(\bL[t],\bq[t+1],\ba[t+1]) | \mathcal{F}_t ] -
\frac{1}{t} \sum_{\tau=1}^t \ell_{\tau}(\bL[t]) \right)\nonumber\\
\stackrel{(a)}{=}{} & \frac{1}{t+1} \left( \mathbb{E} [\ell_1(\bL[t]) ] -
\frac{1}{t} \sum_{\tau=1}^t \ell_{\tau}(\bL[t])\right)\nonumber\\
\leq{} & \frac{1}{t+1} \sup_{\bL[t] \in \mathcal{L}}\left( \mathbb{E}
[\ell_1(\bL[t]) ] - \frac{1}{t} \sum_{\tau=1}^t \ell_{\tau}(\bL[t])\right)
\label{eq:bnd_exp_variations}
\end{align}
where (a) follows from a1). Using the fact that $\ell_i(\bL_t)$ is Lipschitz
from Lemma~\ref{lem:lemma_2}, and uniformly bounded due to a2), Donsker's
Theorem~\cite[Ch. 19.2]{vandervaart_book} yields
\begin{align}
\mathbb{E} \left[ \sup_{\bL[t]} \big| \mathbb{E} [\ell_1(\bL[t]) ] -
\frac{1}{t} \sum_{\tau=1}^t \ell_{\tau}(\bL[t]) \big| \right] =
\mathcal{O}(1/\sqrt{t}). \label{eq:exp_sup_order}
\end{align}
From \eqref{eq:bnd_exp_variations} and \eqref{eq:exp_sup_order} 
the expected non-negative variations can be readily bounded as
\begin{align}
\mathbb{E} \left[ \mathbb{E} \left[ \hat{C}_{t+1}(\bL[t+1]) -
\hat{C}_{t}(\bL[t]) |
\mathcal{F}_t \right]_+ \right] = \mathcal{O}(1/t^{3/2})
\label{eq:bnd_order_exp_variations}
\end{align}
and consequently
\begin{align}
\sum_{t=1}^{\infty} \mathbb{E} \left[ \mathbb{E} \left[
\hat{C}_{t+1}(\bL[t+1]) - \hat{C}_{t}(\bL[t]) |
\mathcal{F}_t \right]_+ \right] < \infty \label{eq:bnd_exp_pos_variations}
\end{align}
which indeed proves that $\{\hat{C}_{t}(\bL[t])\}_{t=1}^\infty$ is a quasi-martingale sequence.

To prove the second part, define first
$U_t(\bL[t]):=C_t(\bL[t])-\frac{\lambda_{\ast}}{2t} \|\bL[t]\|_F^2$ and
$\hat{U}_t(\bL[t]):=\hat{C}_t(\bL[t])-\frac{\lambda_{\ast}}{2t} \|\bL[t]\|_F^2$ for
which $U_t(\bL[t])-\hat{U}_t(\bL[t])=C_t(\bL[t])-\hat{C}_t(\bL[t])$ holds. Following
similar arguments as with $\hat{C}_t(\bL[t])$, one can show
that~\eqref{eq:bnd_exp_pos_variations} holds for $\hat{U}_t(\bL[t])$ as well. It is
also useful to expand the variations
\begin{align}
\hat{U}_{t+1}(\bL[t+1]) - \hat{U}_t(\bL[t]) = \hat{U}_{t+1}(\bL[t+1]) -
\hat{U}_{t+1}(\bL[t]) + \frac{\ell_{t+1}(\bL[t]) - U_t(\bL[t])}{t+1} +
\frac{U_t(\bL[t]) - \hat{U}_t(\bL[t])}{t+1} \nonumber
\end{align}
and bound their expectation conditioned on
$\mathcal{F}_t$, to arrive at
\begin{align}
\frac{U_t(\bL[t])-\hat{U}_t(\bL[t])}{t+1} \leq{} & \left|\mathbb{E} \left[\hat{U}_{t+1}(\bL[t+1])-
\hat{U}_{t+1}(\bL[t])| \mathcal{F}_t\right]\right| + \left|\mathbb{E} \left[\hat{U}_{t+1}(\bL[t+1]) -
\hat{U}_{t}(\bL[t]) | \mathcal{F}_t\right]\right|\nonumber\\
&  + \frac{1}{t+1}
\left|\mathbb{E}[\ell_1(\bL[t])] - \frac{1}{t} \sum_{\tau=1}^t \ell_\tau(\bL[t])
\right|. \label{eq:bnd_Ut_hatUt}
\end{align}
Focusing on the right-hand side of \eqref{eq:bnd_Ut_hatUt}, the second
and third terms are both $\mathcal{O}(1/t^{3/2})$ since counterparts of
\eqref{eq:exp_sup_order} and \eqref{eq:bnd_order_exp_variations} 
also hold for $\hat{U}_t(\bL[t])$. With regards to the first term,
using the fact that $\hat{C}_{t+1}(\bL[t+1]) < \hat{C}_{t+1}(\bL[t])$,
from Lemma~\ref{lem:lemma_2} and a4), it follows that 
$\hat{U}_{t+1}(\bL[t+1]) - \hat{U}_{t+1}(\bL[t]) = o(1/t).$
All in all,
\begin{align}
\sum_{t=1}^{\infty} \frac{\hat{U}_t(\bL[t]) - U_t(\bL[t])}{t+1} < \infty \quad \textrm{a.s.}
\label{eq:bnd_sum_var_hatUt_Ut}
\end{align}
Defining $d_t(\bL[t]):=\hat{U}_t(\bL[t])-U_t(\bL[t])$, due to
Lipschitz continuity of $\ell_t$ and $g_t$ (cf. Lemma~\ref{lem:lemma_2}),
and uniform boundedness of $\{\bL_t\}_{t=1}^{\infty}$ [cf a3)], 
invoking Lemma \ref{lem:lemma_3} one can establish that
$d_{t+1}(\bL[t+1])-d_{t}(\bL[t])=\mathcal{O}(1/t)$.
Hence, Dirichlet's
theorem~\cite{Rudin} applied to the sum~\eqref{eq:bnd_sum_var_hatUt_Ut} asserts that
$\lim_{t \rightarrow \infty} d_t(\bL[t])=0$ a.s., and consequently
$\lim_{t\rightarrow \infty} (\hat{C}_t(\bL[t])-C_t(\bL[t]))=0$ a.s.\hfill$\blacksquare$


\bibliographystyle{IEEEtranS}
\bibliography{IEEEabrv,biblio}

\begin{thebibliography}{10}
\providecommand{\url}[1]{#1}
\csname url@samestyle\endcsname
\providecommand{\newblock}{\relax}
\providecommand{\bibinfo}[2]{#2}
\providecommand{\BIBentrySTDinterwordspacing}{\spaceskip=0pt\relax}
\providecommand{\BIBentryALTinterwordstretchfactor}{4}
\providecommand{\BIBentryALTinterwordspacing}{\spaceskip=\fontdimen2\font plus
\BIBentryALTinterwordstretchfactor\fontdimen3\font minus
  \fontdimen4\font\relax}
\providecommand{\BIBforeignlanguage}[2]{{%
\expandafter\ifx\csname l@#1\endcsname\relax
\typeout{** WARNING: IEEEtranS.bst: No hyphenation pattern has been}%
\typeout{** loaded for the language `#1'. Using the pattern for}%
\typeout{** the default language instead.}%
\else
\language=\csname l@#1\endcsname
\fi
#2}}
\providecommand{\BIBdecl}{\relax}
\BIBdecl

\bibitem{Internet2}
\BIBentryALTinterwordspacing
 [Online]. Available:
  \url{http://internet2.edu/observatory/archive/data-collections.html}
\BIBentrySTDinterwordspacing

\bibitem{ajwak10}
A.~Abdelkefi1, Y.~Jiang, W.~Wang, A.~Aslebo, and O.~Kvittem, ``Robust traffic
  anomaly detection with principal component pursuit,'' in \emph{Proc. of the
  ACM CoNEXT Student Workshop}, Philadelphia, USA, Nov. 2010.

\bibitem{multivariate_onlineanomaly_lakhina07}
T.~Ahmed, M.~Coates, and A.~Lakhina, ``Multivariate online anomaly detection
  using kernel recursive least squares,'' in \emph{Proc. of IEEE/ACM
  International Conference on Computer Communications}, Anchorage, Alaska, May
  2007.

\bibitem{onlinetracking_bolzano10}
L.~Balzano, R.~Nowak, and B.~Recht, ``Online identification and tracking of
  subspaces from highly incomplete information,'' in \emph{Proc. of Allerton
  Conference on Communication, Control, and Computing}, Monticello, USA, Jun.
  2010.

\bibitem{fista}
A.~Beck and M.~Teboulle, ``A fast iterative shrinkage-thresholding algorithm
  for linear inverse problems,'' \emph{SIAM J. Imag. Sci.}, vol.~2, p.
  183–202, Jan. 2009.

\bibitem{Bers}
D.~P. Bertsekas, \emph{Nonlinear Programming}, 2nd~ed.\hskip 1em plus 0.5em
  minus 0.4em\relax Athena-Scientific, 1999.

\bibitem{CCS08}
J.~F. Cai, E.~J. Candes, and Z.~Shen, ``A singular value thresholding algorithm
  for matrix completion,'' \emph{SIAM J. Optim.}, vol.~20, no.~4, pp.
  1956--1982, 2008.

\bibitem{CLMW09}
E.~J. Candes, X.~Li, Y.~Ma, and J.~Wright, ``Robust principal component
  analysis?'' \emph{Journal of the ACM}, vol.~58, no.~1, pp. 1--37, 2011.

\bibitem{CR08}
E.~J. Candes and B.~Recht, ``Exact matrix completion via convex optimization,''
  \emph{Found. Comput. Math.}, vol.~9, no.~6, pp. 717--722, 2009.

\bibitem{CT05}
E.~J. Candes and T.~Tao, ``Decoding by linear programming,'' \emph{IEEE Trans.
  Info. Theory}, vol.~51, no.~12, pp. 4203--4215, 2005.

\bibitem{candes_moisy_mc}
E.~Candes and Y.~Plan, ``Matrix completion with noise,'' \emph{Proceedings of
  the IEEE}, vol.~98, pp. 925--936, 2009.

\bibitem{candes_tutorial}
E.~Candes and M.~Wakin, ``An introduction to compressive sampling,'' \emph{IEEE
  Signal Processing Magazine}, vol.~25, p. 14–20, 2008.

\bibitem{CSPW11}
V.~Chandrasekaran, S.~Sanghavi, P.~R. Parrilo, and A.~S. Willsky,
  ``Rank-sparsity incoherence for matrix decomposition,'' \emph{SIAM J.
  Optim.}, vol.~21, no.~2, pp. 572--596, 2011.

\bibitem{Vaswani_Allerton_11}
Q.~Chenlu and N.~Vaswani, ``Recursive sparse recovery in large but correlated
  noise,'' in \emph{Proc. of 49th Allerton Conf. on Communication, Control, and
  Computing}, Sep. 2011, pp. 752 --759.

\bibitem{petrels_chi12}
Y.~Chi, Y.~C. Eldar, and R.~Calderbank, ``Petrels: Subspace estimation and
  tracking from partial observations,'' in \emph{Proc. of IEEE International
  Conference on Acoustics, Speech and Signal Processing}, Kyoto, Japan, Mar.
  2012.

\bibitem{rank_NP_Duro}
A.~Chistov and D.~Grigorev, ``Complexity of quantifier elimination in the
  theory of algebraically closed fields,'' in \emph{Math. Found. of Computer
  Science}, ser. Lecture Notes in Computer Science.\hskip 1em plus 0.5em minus
  0.4em\relax Springer Berlin / Heidelberg, 1984, vol. 176, pp. 17--31.

\bibitem{elements_of_statistics}
T.~Hastie, R.~Tibshirani, and J.~Friedman, \emph{The Elements of Statistical
  Learning}, 2nd~ed.\hskip 1em plus 0.5em minus 0.4em\relax Springer, 2009.

\bibitem{increm_grad_grasmannian_bolzano12}
J.~He, L.~Balzano, and A.~Szlam, ``Incremental gradient on the {G}rassmannian
  for online foreground and background separation in subsampled video,'' in
  \emph{Proc. of IEEE Conference on Computer Vision and Pattern Recognition},
  Providence, Rhode Island, Jun. 2012.

\bibitem{moura_dnesterov}
D.~Jakovetic, J.~Xavier, and J.~M.~F. Moura, ``Fast distributed gradient
  methods,'' arXiv:1112.2972v1 [cs.IT].

\bibitem{LCD04}
A.~Lakhina, M.~Crovella, and C.~Diot, ``Diagnosing network-wide traffic
  anomalies,'' in \emph{Proc. of ACM SIGCOMM}, Portland, OR, Aug. 2004.

\bibitem{LPC04}
A.~Lakhina, K.~Papagiannaki, M.~Crovella, C.~Diot, E.~D. Kolaczyk, and N.~Taft,
  ``Structural analysis of network traffic flows,'' in \emph{Proc. of ACM
  SIGMETRICS}, New York, NY, Jul. 2004.

\bibitem{Ljung}
L.~Ljung and T.~S$\ddot{o}$derstr$\ddot{o}$m, \emph{Theory and Practice of
  Recursive Identification}, 2nd~ed.\hskip 1em plus 0.5em minus 0.4em\relax MIT
  Press, 1983.

\bibitem{mairalonlinelearning}
J.~Mairal, J.~Bach, J.~Ponce, and G.~Sapiro, ``Online learning for matrix
  factorization and sparse coding,'' \emph{J. of Machine Learning Research},
  vol.~11, pp. 19--60, Jan. 2010.

\bibitem{tsp_rankminimization_2012}
M.~Mardani, G.~Mateos, and G.~B. Giannakis, ``In-network sparsity regularized
  rank minimization: Applications and algorithms,'' \emph{IEEE Trans. Signal
  Process.}, Feb. 2012 (submitted), see also arXiv:1203.1570v1 [cs.MA].

\bibitem{tit_exactrecovery_2012}
------, ``Recovery of low-rank plus compressed sparse matrices with application
  to unveiling traffic anomalies,'' \emph{IEEE Trans. Info. Theory.}, Apr. 2012
  (submitted), see also arXiv:1204.6537v1 [cs.IT].

\bibitem{gonzalo_rpca}
G.~Mateos and G.~B. Giannakis, ``Robust pca as bilinear decomposition with
  outlier-sparsity regularization,'' \emph{IEEE Trans. Signal Process.}, Sep.
  2012, see also arXiv:1111.1788v1 [stat.ML].

\bibitem{distributed_pca_heor12}
Z.~Meng, A.~Wiesel, and A.~Hero, ``Distributed principal component analysis on
  networks via directed graphical models,'' in \emph{Proc. of IEEE
  International Conference on Acoustics, Speech, and Signal Processing}, Kyoto,
  Japan, Mar. 2012.

\bibitem{Natarajan_NP_duro}
B.~K. Natarajan, ``Sparse approximate solutions to linear systems,'' \emph{SIAM
  J. Comput.}, vol.~24, pp. 227--234, 1995.

\bibitem{nesterov83}
Y.~Nesterov, ``A method of solving a convex programming problem with
  convergence rate $o(1/k^2)$,'' \emph{Soviet Mathematics Doklady}, vol.~27,
  pp. 372--376, 1983.

\bibitem{RFP07}
B.~Recht, M.~Fazel, and P.~A. Parrilo, ``Guaranteed minimum-rank solutions of
  linear matrix equations via nuclear norm minimization,'' \emph{SIAM Rev.},
  vol.~52, no.~3, pp. 471--501, 2010.

\bibitem{Recht_Parallel_2011}
B.~Recht and C.~Re, ``Parallel stochastic gradient algorithms for large-scale
  matrix completion,'' 2011, (submitted).

\bibitem{Roughan}
M.~Roughan, ``A case study of the accuracy of {SNMP} measurements,''
  \emph{Journal of Electrical and Computer Engineering}, Dec. 2010, article ID
  812979.

\bibitem{Rudin}
W.~Rudin, \emph{Principles of Mathematical Analysis}, 3rd~ed.\hskip 1em plus
  0.5em minus 0.4em\relax McGraw-Hill, 1976.

\bibitem{Solo_Adaptive_Book}
V.~Solo and X.~Kong, \emph{Adaptive Signal Processing Algorithms: Stability and
  Performance}.\hskip 1em plus 0.5em minus 0.4em\relax Prentice Hall, 1995.

\bibitem{MC03}
M.~Thottan and C.~Ji, ``Anomaly detection in {IP} networks,'' \emph{IEEE Trans.
  Signal Process.}, vol.~51, pp. 2191--2204, Aug. 2003.

\bibitem{tropp06tit}
J.~Tropp, ``Just relax: Convex programming methods for identifying sparse
  signals,'' \emph{IEEE Trans. Info. Theory}, vol.~51, pp. 1030--1051, Mar.
  2006.

\bibitem{tseng_cnvg_bcd}
P.~Tseng, ``Convergence of a block coordinate descent method for
  nondifferentiable minimization,'' \emph{Journal of optimization theory and
  applications}, vol. 109, pp. 475--494, 2001.

\bibitem{vandervaart_book}
A.~W. {Van Der Vaart}, \emph{Asymptotic Statistics}.\hskip 1em plus 0.5em minus
  0.4em\relax Cambridge University Press, 2000.

\bibitem{yang95}
B.~Yang, ``Projection approximation subspace tracking,'' \emph{IEEE Trans.
  Signal. Process.}, vol.~43, pp. 95--107, Jan. 1995.

\bibitem{zggr05}
Y.~Zhang, Z.~Ge, A.~Greenberg, and M.~Roughan, ``Network anomography,'' in
  \emph{Proc. of ACM SIGCOM Conf. on Interent Measurements}, Berekly, CA, USA,
  Oct. 2005.

\bibitem{zrwq09}
Y.~Zhang, M.~Roughan, W.~Willinger, and L.~Qiu, ``Spatio-temporal compressive
  sensing and internet traffic matrices,'' in \emph{Proc. of ACM SIGCOM Conf.
  on Data Commun.}, New York, USA, Oct. 2009.

\bibitem{zlwcm10}
Z.~Zhou, X.~Li, J.~Wright, E.~Candes, and Y.~Ma, ``Stable principal component
  pursuit,'' in \emph{Proc. of Intl. Symp. on Information Theory}, Austin, TX,
  Jun. 2010, pp. 1518--1522.

\end{thebibliography}

\end{document}